\documentclass[a4paper,10pt,unnumberedsections,twoside]{LTJournalArticle}

\usepackage[backend=biber]{biblatex}

\usepackage[table]{xcolor}
\usepackage{amsmath}
\usepackage{amsfonts}
\usepackage{amssymb}
\usepackage{geometry}
\usepackage{graphicx}
\usepackage{wrapfig}
\usepackage{fancyhdr}
\usepackage{setspace}
\usepackage{booktabs}
\usepackage{enumitem}
\usepackage{multirow}
\usepackage{tcolorbox}
\usepackage{bm}

\usepackage{blindtext}
\usepackage{titlesec}
\usepackage[normalem]{ulem}
\usepackage{float}
\usepackage{caption}

\newlist{checkbox}{itemize}{2}
\setlist[checkbox]{label=$\square$}

\titleformat{\paragraph}
{\normalfont\normalsize\bfseries}{\theparagraph}{1em}{}
\titlespacing*{\paragraph}
{0pt}{3.25ex plus 1ex minus .2ex}{1.5ex plus .2ex}

\title{When Transformers Meet Recommenders: Integrating Self-Attentive Sequential Recommendation with Fine-Tuned LLMs} % Article title, use manual lines breaks (\\) to beautify the layout

\author{%
	Kechen Liu\textsuperscript{1}
}

\date{\footnotesize\textsuperscript{\textbf{1}} BSc Computer Science with Industrial Placement, University of Exeter, kl572@exeter.ac.uk \\}

\renewcommand{\maketitlehookd}{%
	\begin{abstract}
		\noindent Self-Attentive Sequential Recommendation (SASRec) effectively captures long-term user preferences by applying attention mechanisms to historical interactions. Concurrently, the rise of Large Language Models (LLMs) has motivated research into LLM-based recommendation, which leverages their powerful generalization and language understanding capabilities. However, LLMs often lack the domain-specific knowledge and collaborative signals essential for high-quality recommendations when relying solely on textual prompts. To address this limitation, this study proposes SASRecLLM, a novel framework that integrates SASRec as a collaborative encoder with an LLM fine-tuned using Low-Rank Adaptation (LoRA). The components are connected via a mapping layer to align their dimensional spaces, and three targeted training strategies are designed to optimize the hybrid architecture. Extensive experiments on multiple datasets demonstrate that SASRecLLM achieves robust and consistent improvements over strong baselines in both cold-start and warm-start scenarios. This work advances the field of LLM-based recommendation by presenting a modular and effective paradigm for fusing structured collaborative filtering with the semantic power of fine-tuned LLMs. The implementation is available on GitHub: \nolinkurl{https://github.com/kechenkristin/RecLLM}.
	\end{abstract}
}

\begin{document}

\maketitle % Output the title section

%----------------------------------------------------------------------------------------
%	ARTICLE CONTENTS
%----------------------------------------------------------------------------------------

\section{Introduction}

Large Language Models (LLMs), such as LLaMA \cite{touvron2023llama} and ChatGPT \cite{ouyang2022training}, have seen rapid advancements recently. Previous studies highlight their strong generalizability \cite{ouyang2022training}, enabling them to integrate and apply knowledge in diverse domains. Additionally, they excel in Natural Language Understanding (NLU), making them highly effective for information retrieval and reasoning tasks \cite{naveed2023comprehensive}.

Recommender Systems (RSs) play a critical role in information filtering by delivering personalized content to users. They typically rely on collaborative information, which refers to patterns in user-item interactions to infer preferences across a user base \cite{burke2011recommender}. At the core of most RSs are Conventional Recommendation Models (CRMs), which include methods such as Matrix Factorization (MF) and Neural Collaborative Filtering (NCF). Within this category, Sequential Recommender Systems (SRSs) extend CRMs by modeling the temporal order of user interactions to predict future behavior more effectively \cite{kang2018sasrec}. Each item is assigned to a learnable embedding optimized to model user preferences and capture sequential patterns from past interactions \cite{liao2024llara}. Recent advances in deep learning have popularized Self-Attentive Sequential Recommendation (SASRec) for its enhanced ability to model complex user behavior patterns \cite{sun2019bert4rec} and efficiently capture long-range dependencies in user interaction sequences \cite{zhou2020selfsupervised} with self-attentive mechanisms, allowing adaptive focus on relevant past interactions when predicting future engagement \cite{zhou2020s3rec}.

Although CRMs effectively capture user–item interactions, they lack the ability to incorporate high-level semantics and contextual reasoning. This limitation has led to growing interest in LLMs for Recommendation (LLM4Rec), which leverage natural language understanding (NLU) and generalization capabilities to a enable more flexible, context-aware predictions \cite{zhao2023recommender}. A common strategy for adapting LLMs to recommendation tasks is In-Context Learning (ICL), which prompts LLMs with examples without retraining. However, relying solely on ICL often results in poor performance, as LLMs are not specifically trained for recommendation objectives \cite{bao2023tallrec}. To address this, fine-tuning is necessary to adapt LLMs to the recommendation objectives. However, despite fine-tuning, LLMs still struggle to capture collaborative signals that are critical to accurate recommendations. Therefore, this study introduces SASRecLLM, a novel framework that combines the strengths of both methods. It builds a SASRec encoder to extract collaborative information from user-item interactions, fine-tunes the LLM with lightweight LoRA to generate final recommendations, and uses a mapping layer to align collaborative embeddings with the LLM’s semantic space. By combining SASRec with LLM, the proposed hybrid design effectively encodes collaborative signals in LLM4Rec, enhancing accuracy across both warm and cold start settings while allowing personalized cross-domain recommendations. 

%------------------------------------------------

\section{Related Work}

\subsection{Recommender Systems} \label{sec: RSs}

With the exponential growth of digital content, RSs play a critical role in capturing collaborative signals, which refer to data derived from user interactions, preferences, and behaviors. These signals are used to generate personalized recommendations \cite{burke2011recommender}. RSs typically operate under three tasks: \textbf{ (i) binary classification}, predicting whether a user will engage with an item (Yes/No) \cite{aggarwal2016collaborative}; \textbf{(ii) Top-K Ranking}, identifying and ranking the most relevant items \cite{wang2020sequential}; and \textbf{ (iii) multiclass classification}, predicting different types of user interactions \cite{ricci2015recsys}.

CRMs serve as the foundation for modern RSs. Collaborative Filtering (CF), a widely used CRM, predicts user preferences by identifying similarities between users or items based on past interactions \cite{ricci2015recsys}. However, CF struggles to adapt to evolving user preferences due to its reliance on static user-item interactions. SRSs overcome this limitation by leveraging the temporal sequence of user actions, enabling more dynamic recommendations \cite{wang2020sequential}. Early SRSs models, such as Markov Chains (MCs), predict user actions using probability transition matrices \cite{ahmed2016markov} but fail to capture long-range dependencies. Recurrent Neural Networks (RNNs) improve upon this by incorporating memory mechanisms, but they suffer from vanishing gradients and are computationally inefficient for long user histories \cite{hidasi2016session}. Recently, the Transformer model has redefined deep learning with self-attention mechanisms, enabling models to selectively focus on relevant past interactions instead of processing sequences in a fixed order \cite{vaswani2017attention}. This mechanism improves interpretability by dynamically weighting past interactions on the basis of their importance, ensuring that relevant historical behaviors influence future predictions while filtering out less relevant signals. SASRec, a Transformer-based SRS, builds on this idea by effectively capturing long-term dependencies such as RNNs while also maintaining the flexibility of MCs to make predictions based on a small number of recent actions \cite{kang2018sasrec}. This balance allows SASRec to model user behavior sequences holistically while adapting to localized interactions. Furthermore, by processing all previous interactions in parallel, SASRec improves computational efficiency and improves its ability to model complex relationships between user actions, even when they are not adjacent \cite{sun2019bert4rec, desouza2021transformers4rec}.

\subsection{Large Language Models}
Large Language Models (LLMs) are neural networks trained in vast text corpora to understand and generate human-like language \cite{naveed2023comprehensive}. The model's input is a sequence of text, formally known as a "prompt," which is first converted into a numerical format through a process called tokenization. The model's output is then generated autoregressively; it predicts the subsequent text one token (a word or sub-word) at a time, with each new token being conditioned on the sequence of all preceding ones. This process allows the generated text to be a contextually relevant continuation, answer, or creative expansion of the original prompt.  Unlike CRMs, which rely on structured data, LLMs leverage extensive world knowledge to interpret user intent and generate personalized recommendations, enhancing accuracy and engagement. Moreover, their ability to dynamically adjust outputs based on real-time input improves adaptability \cite{yu2025application}. Furthermore, LLMs excel at ICL. This includes zero-shot learning, the ability to perform tasks based on instructions alone without any prior examples, and few-shot learning, where performance is improved by including a small number of examples directly within the input prompt. These capabilities are crucial for recommendation systems, as they allow the model to infer user preferences and mitigate cold-start challenges using minimal task-specific data \cite{sanner2023large}. This section explores their training mechanisms, limitations in specialization, and fine-tuning techniques for optimizing recommendation quality.

\begin{figure} % Single column figure
\includegraphics[width=\linewidth]{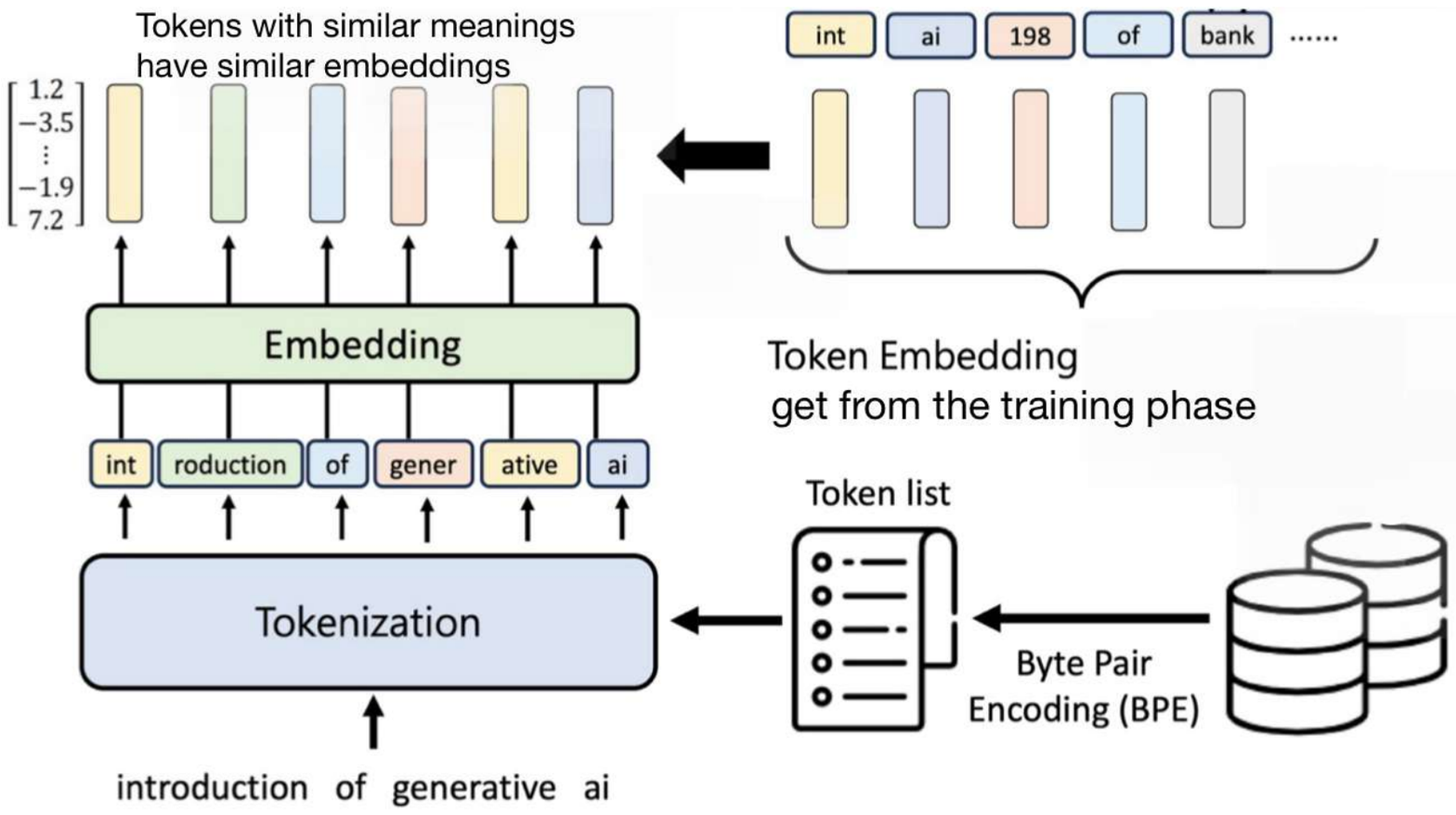}
	\caption{LLMs tokenize input prompts, embed them, and perform inference on the embeddings. Adapted from \cite{lee2024transformerslide}.}
    \label{fig: transformer}
\end{figure}

As depicted in Fig. \ref{fig: transformer}, LLMs' training mechanisms enable them to process complex, information-rich inputs through two key phases: 

\textbf{(i) Training and Tokenization}: During training, LLMs learn from vast datasets, establishing a comprehensive knowledge base \cite{yang2024harnessing}. When processing real-world prompts, input text is first segmented into tokens. These tokens are then mapped to high-dimensional vectors via embedding layers, preserving semantic relationships in the embedding space \cite{wu2024towards}. Transformer architectures and self-attention mechanisms further enhance contextual understanding, improving token associations \cite{vaswani2017attention}. 

\textbf{(ii) Inference}: Inference allows LLMs to dynamically apply learned knowledge across various text-based tasks, tailoring responses to real-time inputs \cite{zhao2023recommender}. A multi-layer perceptron (MLP) processes token representations within the Transformer’s feed-forward layers, refining contextual embeddings before final token prediction \cite{liu2018neural, ozdemir2023quick}.

Pre-trained LLMs generalize well across domains but often produce generic responses rather than task-specific ones. Fine-tuning addresses this limitation by training the model on domain-specific data, optimizing its parameters for specialization \cite{naveed2023comprehensive}. The most direct method, full fine-tuning, updates all parameters via backpropagation and gradient descent. However, it requires significant computational resources and labeled data, making it impractical in many cases. As LLMs scale, full fine-tuning becomes increasingly costly and prone to overfitting, particularly with small datasets \cite{lv2023full}. To address these challenges, Parameter-Efficient Fine-Tuning (PEFT) optimizes only a small subset of parameters, reducing computational costs and improving efficiency. Among PEFT methods, Low-Rank Adaptation (LoRA) is particularly efficient. Unlike adapters, which add new modules, or prefix tuning, which prepends virtual tokens, LoRA directly integrates low-rank matrices into existing layers, reducing memory consumption. During training, LoRA adjusts model weights through two low-rank matrices while keeping the original parameters frozen. After training, these matrices merge with the existing model, ensuring LoRA maintains inference speed, unlike Adapter-based methods that introduce additional layers and increase latency \cite{hu2022lora}.

\subsection{LLMs for Recommendations}\label{sec: llm4rec}
Despite recent advancements, CRMs still face critical limitations that hinder their real-world effectiveness. While CRMs perform well when historical interactions are rich, they often struggle in cold-start scenarios, which occur when there is insufficient interaction data to learn user preferences or assess item relevance \cite{gogna2015comprehensive}. Moreover, solely relying on collaborative modeling leads to limited explainability and an incomplete understanding of evolving user intent in RSs \cite{nunes2017systematic}. Finally, they are domain-specific and operate primarily on discrete ID-based features, limiting their capacity to interpret rich content or transfer user preference knowledge across domains \cite{islam2022systematic}. In parallel, recent progress in LLMs has given rise to the emergence of LLMs for Recommendations (LLM4Rec), a research direction that explores integrating LLMs into RSs to enhance recommendation quality. However, LLM4Rec faces two key challenges: \textbf{(i) whether to fine-tune LLMs during training} and \textbf{(ii) whether to incorporate CRMs during inference} \cite{lin2023can}.

% \begin{wrapfigure}{r}{0.7\textwidth} % 'r' = right side, 60% of text width
%     \centering
%     \vspace{-10pt}
%     \includegraphics[width=\linewidth]{img/ICL_fail.pdf}
%     \caption{Illustration of ChatGPT refusing to answer or always predicting positively. Adapted from \cite{bao2023tallrec}.}
%     \label{fig:ICL}
%     % \vspace{-10pt}
% \end{wrapfigure}

\begin{figure*} % Two column figure (notice the starred environment)
	\includegraphics[width=\linewidth]{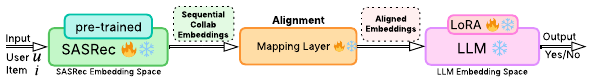}
\caption{\textbf{Architecture of SASRecLLM.} The model takes user–item interactions as input, extracts collaborative signals via \textbf{SASRec}, projects them into the \textbf{LLM} embedding space through a \textbf{mapping layer}, and generates predictions using a fine-tuned LLM enhanced with \textbf{LoRA}. }
%     \label{fig: architecture}
    \label{fig: architecture}
\end{figure*}

ICL enables LLM4Rec by utilizing natural language prompts without explicit fine-tuning or CRMs. This approach enhances adaptability and mitigates cold-start issues through contextual reasoning \cite{brown2020language}. However, relying solely on ICL can hinder recommendation accuracy, as research shows ChatGPT often fails to respond or defaults to positive predictions (e.g., ‘likes’). This limitation likely stems from the lack of task-specific training \cite{bao2023tallrec}. To address this challenge, researchers proposed TALLRec, an innovative LLM4Rec framework that applies fine-tuning to enhance recommendation performance \cite{bao2023tallrec}. When excluding CRMs, this approach reformulates recommendation problems (e.g., click-through rate estimation, next-item prediction) as text classification or sequence-to-sequence tasks. By updating model parameters or applying prompt engineering techniques based on user preferences and item interactions, fine-tuning enhances recommendation relevance, making it particularly effective in cold-start settings \cite{boz2024improving}. Scalability still remains a challenge. While early studies applied full fine-tuning to small-scale models (e.g., BERT-base, LongFormer) \cite{lin2023can}, the increasing size of modern LLMs makes this approach computationally prohibitive, necessitating more efficient fine-tuning strategies. Moreover, in CRMs, collaborative information serves as a unique interaction signal, capturing user-item co-occurrence relationships within engagement data \cite{zhang2023collm}. Especially in warm start scenarios, where historical user-item interactions offer valuable behavioral cues, LLMs often lack structured mechanisms to process these signals effectively. As a result, they may struggle to capture long-term engagement patterns, which can reduce recommendation accuracy. \cite{wang2024llm}. To this end, research suggests integrating CRMs into fine-tuned LLMs to leverage collaborative knowledge alongside language-based reasoning \cite{zheng2024adapting}. Early attempt (2021–2022) primarily focused on fine-tuning smaller pre-trained models for text-rich recommendation tasks, such as news recommendation and web search \cite{wu2021empowering}. In these settings, CRMs generated the final recommendation, while pre-trained models enriched input representations. Recently, with the emergence of billion-parameter models such as ChatGPT and LLaMA (2023), researchers have begun treating LLMs and CRMs equally rather than making LLMs passive feature encoders \cite{lin2023can}. 

Motivated by this, this project explores a promising direction by integrating CRMs as independent collaborative modeling encoder, merging the world knowledge, reasoning, and instruction-following capabilities of LLMs with structured user-item interaction modeling. Furthermore, the use of lightweight fine-tuning via LoRA enhances the LLM’s ability to adapt to domain-specific recommendation tasks.

%------------------------------------------------

\section{Methodology}

SASRecLLM is designed as a hybrid model that integrates SASRec encoder with an LLM layer via a mapping layer. Given a user $u$ and item $i$, the expected output \(y \in \{0, 1\}\) indicates whether $u$ likes or dislikes $i$. While SASRec extracts temporal user-item interaction patterns and generates latent representations, the LLM enriches these signals with semantic reasoning (i.e., interpreting item descriptions or user reviews) to generate the final output. To bridge these modalities, a mapping layer projects SASRec’s embeddings into the LLM’s token space, enabling seamless information transfer between the two models. As illustrated in Fig. \ref{fig: architecture}, SASRecLLM consists of three core components:

\noindent \textbf{(i) SASRec Model}: SASRec encodes user and item IDs, capturing interaction signals and learning sequential user behavior to generate sequential collaborative embeddings. 

\noindent \textbf{(ii) Mapping Layer}: Projects SASRec’s collaborative embeddings into the LLM’s token space, enabling modality alignment. 

\noindent \textbf{(iii) LLM}: Generates the final output through semantic reasoning over textual data and leverages world knowledge. An additional LoRA layer is incorporated to fine-tune the LLM for better understanding of the recommendation task. 

Moreover, three training strategies are leveraged to train the model parameters: Dual-Stage Training, Hierarchical Freezing, and Plug-and-Play Tuning.
% TODO: remove this sentence
This section formalizes the problem, details the architecture, and explains the training strategies.

\subsection{Model Architecture}
\subsubsection{SASRec}
In the SASRecLLM architecture, SASRec encodes user and item IDs as latent embeddings to capture collaborative signals. These embeddings, \( u', i' \in \mathbb{R}^{d_1} \), are computed as:
\begin{equation}\label{eq:sasrec}
{u'} = r_{\psi}({u}), \quad
{i'} = r_{\psi}({i}),
\end{equation}
where \( r_{\psi}(\cdot) \) denotes the SASRec transformation process, \( d_1 \) represents the embedding dimensionality for SASRec, and \( \psi \) corresponds to the model parameters.

\begin{figure}[htbp] % Single column figure
    \includegraphics[width=\linewidth]{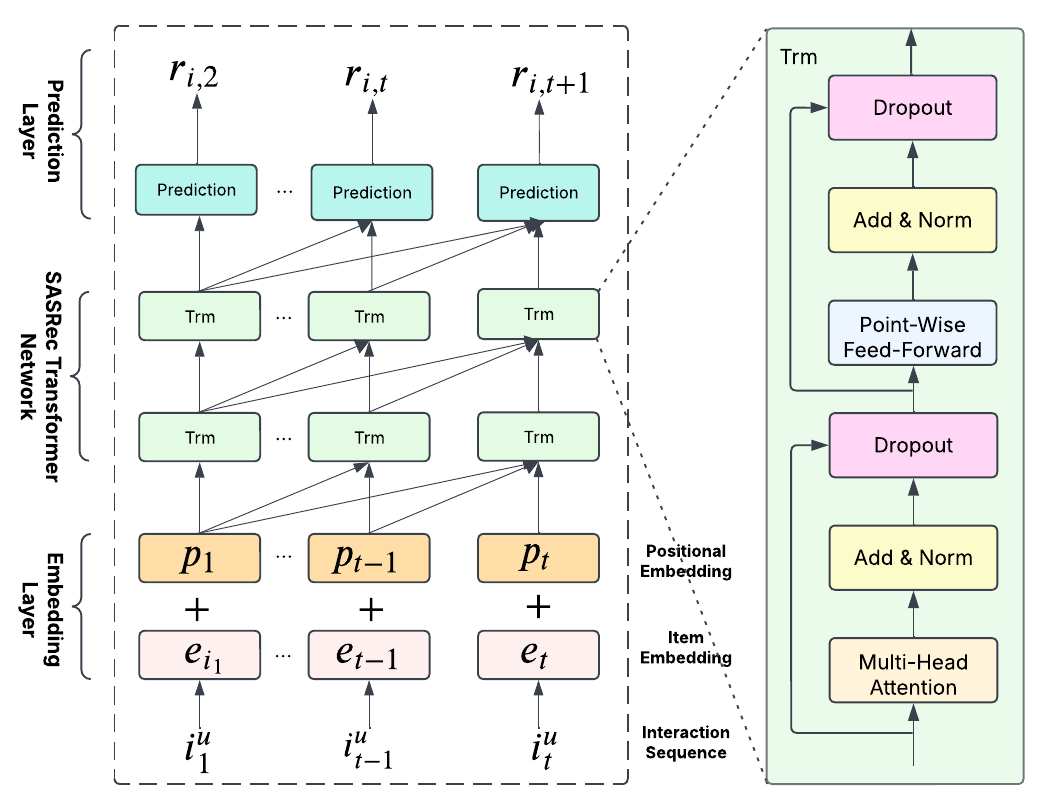}
    \caption{\textbf{Architecture of SASRec.} The model consists of three layers and uses a \textbf{self-attention} mechanism to consider all previous items at each step, focusing on those most relevant to the next action. \textbf{“Trm”} denotes Transformer blocks.}
    \label{fig: SASRec2}
\end{figure}

SASRec is trained to model sequential dependencies by learning patterns from past interactions for future engagement prediction. As a sequence prediction model, it processes user interactions chronologically, leveraging self-attention to identify relevant past actions and capture long-term dependencies. To achieve this, as illustrated in Fig. \ref{fig: SASRec2}, SASRec consists of three key components:  
(i) an \textbf{Embedding Layer} that maps items into a latent space,  
(ii) a \textbf{SASRec Transformer Network} for learning complex item transitions, and  
(iii) a \textbf{Prediction Layer} that computes relevance scores.  

Formally, given a set of users \(\mathcal{U} = \{u_1, u_2, \dots, u_{|\mathcal{U}|}\}\), a set of items \(\mathcal{I} = \{i_1, i_2, \dots, i_{|\mathcal{I}|}\}\), and the interaction sequence \(S^u = \{i_1^u, i_2^u, \dots, i_{n_u}^u\}\), the goal is to predict the next item \( i_{n_u + 1}^u \) that user \( u \) is most likely to interact with. Here, \( n_u \) denotes the number of items interacted with by user \( u \), and \( i_t^u \) (\( 1 \leq t \leq n_u \)) represents the item that user \( u \) interacted with at the position \( t \) in the sequence. The input sequence \(S^u\) is standardized to a fixed-length sequence \(s = \{s_1,s_2,\dots,s_n\}\). Sequences longer than \(n\) retain only the most recent \(n\) interactions, while shorter sequences are left-padded with a placeholder token \(\texttt{<pad>}\) to maintain length \(n\). The target output \(o_t\) at each time step \(t\) is defined as:

\begin{equation}
o_t=\begin{cases}
\texttt{<pad>}									&	\text{if } s_t \text{ is a padding item},\\
s_{t+1}									&	1\leq t<n,\\
i_{n_u}^u		&	t=n.
\end{cases}
\end{equation}

\paragraph{Embedding Layer} 

This layer represents user interactions using an item embedding matrix \(E_{\mathcal{I}} \in \mathbb{R}^{|\mathcal{I}| \times d_1}\) and a positional embedding matrix \(E_{\mathcal{P}} \in \mathbb{R}^{n \times d_1}\), which encodes item sequence positions to compensate for self-attention’s lack of order sensitivity. For each item in the user’s interaction sequence, the item embedding is retrieved as \( e_{i_t} = E_{\mathcal{I}}[ i_t^u ] \), and the positional embedding is \( p_t = E_{\mathcal{P}}[t] \), where \( t \) represents the item's position in the sequence. The resulting embedding matrix \( \widehat{\mathbf{E}} \in \mathbb{R}^{n \times d_1} \) is computed as:
\begin{equation}
\widehat{\mathbf{E}} =
\begin{bmatrix}
    e_{i_1} + p_1 \\
    e_{i_2} + p_2 \\
    \vdots \\
    e_{i_n} + p_n
\end{bmatrix}.
\end{equation}

% ~\\
% \noindent $\bullet$ \textbf{SASRec Transformer Network} 
\paragraph{SASRec Transformer Network} 

\noindent The SASRec Transformer Network consists of multiple stacked SASRec Transformer layers, each enhancing sequence modeling by attending to the most relevant interactions through three key components:  
% (i) A Self-Attention Layer that captures contextual relationships by attending to relevant interactions.  
% (ii) A Point-Wise Feed-Forward Network (FFN) that enhances feature transformation through non-linearity.  
% (iii) A Residual Connection that stabilizes training and improves gradient flow.

\noindent \textbf{(i) Self-Attention Layer}  

\noindent This layer captures contextual relationships by attending to relevant interactions. This is achieved through transforming input embeddings \(\widehat{\mathbf{E}}\) into query (\(Q\)), key (\(K\)), and value (\(V\)) matrices via learnable weights \(W^Q, W^K, W^V \in \mathbb{R}^{d_1 \times d_1}\), and applies scaled dot-product attention \cite{vaswani2017attention} to capture dependencies:  
\begin{equation}
\begin{array}{c}
S = \textbf{SA}(\widehat{\mathbf{E}}) = \text{Attention}\left(\widehat{\mathbf{E}}W^Q, \widehat{\mathbf{E}}W^K, \widehat{\mathbf{E}}W^V\right), \\
\text{Attention}(Q, K, V) = \text{softmax}\left(\frac{QK^\top}{\sqrt{r}}\right)V. 
\end{array}
\end{equation}
Here, each row in \( Q, K, V \) corresponds to an item in the sequence. The softmax operation normalizes attention scores to determine item dependencies, while the scaling factor \(\sqrt{r}\) prevents large values from dominating the attention distribution.

~\\
\noindent \textbf{(ii) Point-Wise Feed-Forward Network}  

\noindent The Self-Attention layer aggregates information across items, but the outputs remain linear. To introduce nonlinearity and enhance feature interactions, a shared two-layer Point-Wise Feed-Forward Network (FFN) is applied to each position \(\mathbf{s}_i\):
\begin{equation} 
\mathbf{F}_i = \text{FFN}(\mathbf{s}_i) = \text{ReLU}\left(\mathbf{s}_i \mathbf{W}^{(1)} + \mathbf{b}^{(1)}\right)\mathbf{W}^{(2)} + \mathbf{b}^{(2)},
% \label{eq:ffn}  
\end{equation} 
where \(\mathbf{W}^{(1)}, \mathbf{W}^{(2)} \in \mathbb{R}^{d_1 \times d_1}\) and \(\mathbf{b}^{(1)}, \mathbf{b}^{(2)} \in \mathbb{R}^{d_1}\) are the weight matrices and bias terms, respectively.

~\\
\noindent \textbf{(iii) Residual Connection}  

\noindent As the number of Transformers in the network increases, a residual connection is applied to enhance stability, prevent overfitting, and reduce training costs. Formally,
\begin{equation}
l(x) = x + \text{Dropout}(l(\text{LayerNorm}(x))),
\end{equation}
where \(l(x)\) represents either the self-attention layer \(\mathbf{SA}\) or the feed-forward network \(\mathbf{F}\). For each layer \(f\) in the network, layer normalization is applied to the input \(x\) for stability and faster training. The processed input is then fed into \(l(x)\), where dropout is applied to mitigate overfitting. Finally, a residual connection adds the original input \(x\) to the output.

% ~\\
% \noindent $\bullet$ \textbf{Prediction Layer} 
\paragraph{Prediction Layer} 
\noindent After $b$ Transformers encode past interactions, the model predicts the next item using ${l_t^{(b)}}$, which encodes the first $t$ processed items. Formally, the relevance score of item i, given the sequence \( (s_1,s_2,\dots,s_t) \), is computed as:  $\mathbf{r_{i,t+1}} = l_t^{(b)} \mathbf{e}_{v_t}^\top$. Higher scores indicate relevance, enabling the model to rank items for recommendations.

\begin{figure} % Single column figure
	\includegraphics[width=\linewidth]{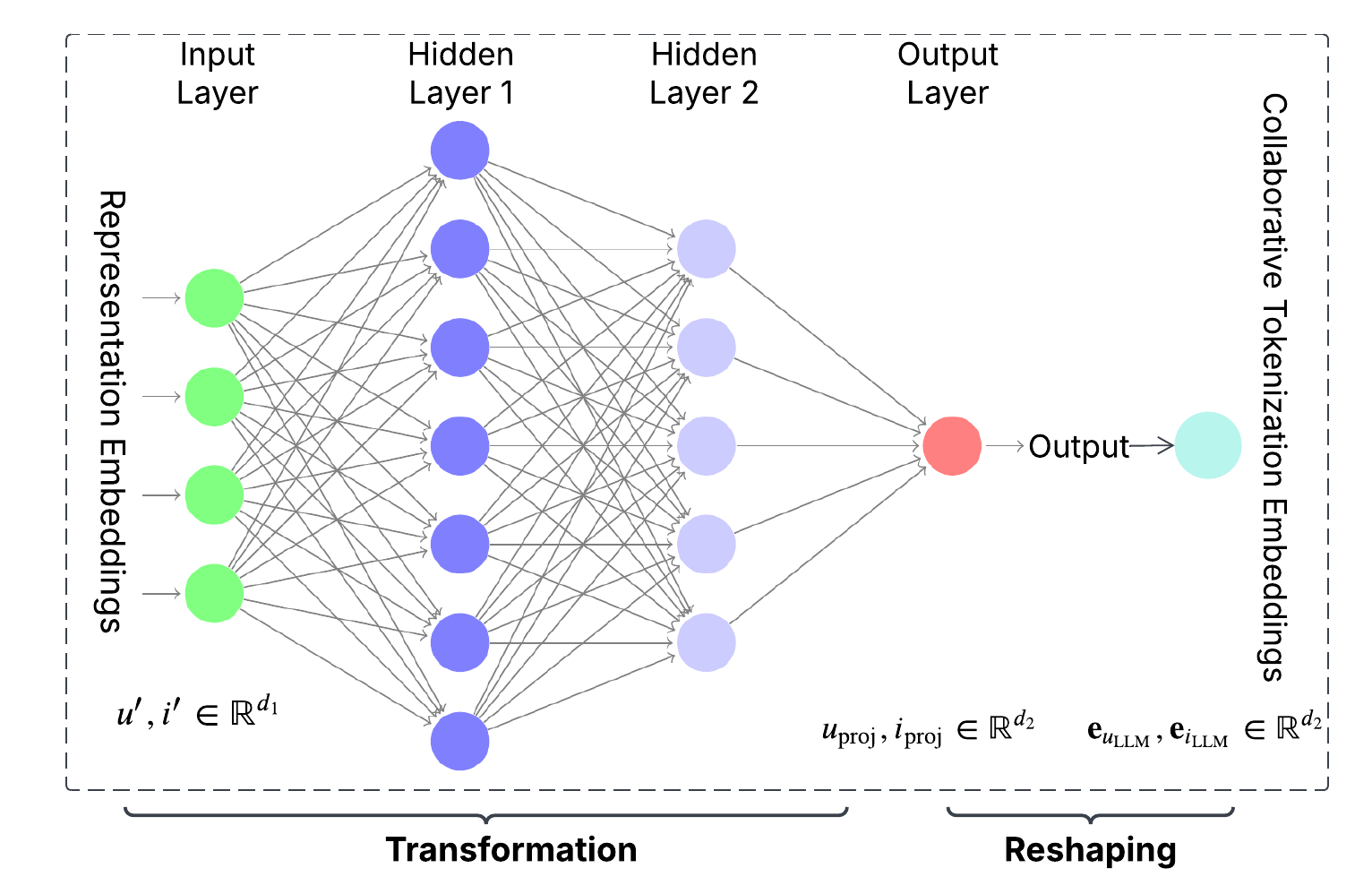}
    \caption{\textbf{Architecture of the mapping layer.} Implemented as a \textbf{MLP} with two hidden layers, it receives embeddings from SASRec and aligns them to the LLM's token embedding space through two phases: \textbf{Transformation} and \textbf{Reshaping}.
}
    \label{fig: MLP2}
\end{figure}

\subsubsection{Mapping Layer}
To align SASRec’s learned representations with the LLM’s token embeddings, a mapping layer is introduced. Implemented as a MLP, this layer transforms collaborative embeddings \( u', i' \in \mathbb{R}^{d_1} \) into the LLM’s token embedding space \( \mathbf{e}_{u_\text{LLM}}, \mathbf{e}_{i_\text{LLM}} \in \mathbb{R}^{d_2} \). Formally,
\begin{equation}\label{eq:mapping_layer}
\mathbf{e}_{u_\text{LLM}} = m_{\phi}({u'}), \quad
\mathbf{e}_{i_\text{LLM}} = m_{\phi}({i'}),
\end{equation}

\noindent where \( m_{\phi}(\cdot) \) represents the mapping process, \( d_1 \) and \( d_2 \) denote the embedding dimensionalities of SASRec and LLM, respectively (\( d_1 \) $<$ \( d_2 \)), and \( \phi \) corresponds to the mapping layer’s parameters. As illustrated in Fig. \ref{fig: MLP2}, the mapping process consists of two key phases: transformation and Reshaping.

% ~\\
% \noindent $\bullet$ \textbf{Transformation} 
\paragraph{Transformation} 
\noindent The collaborative embeddings are first expanded to a higher-dimensional space, allowing the model to learn richer transformations. They are then projected into the LLM’s expected embedding size \(d_2\):
\begin{equation} 
\mathbf{x}_{\text{proj}} = \text{ReLU}\left(\mathbf{x}' \mathbf{W}_p^{(1)} + \mathbf{b}_p^{(1)}\right)\mathbf{W}_p^{(2)} + \mathbf{b}_p^{(2)} \in \mathbb{R}^{d_2},  
% \label{eq:mlp_proj}  
\end{equation}  
where \( \mathbf{x'} \) represents either \( i' \) or \( u' \), \(\mathbf{W}_p^{(1)} \in \mathbb{R}^{d_1 \times d_{\text{exp}}}\) is a learnable weight matrix for expansion, and \(\mathbf{W}_p^{(2)} \in \mathbb{R}^{d_{\text{exp}} \times d_2}\) projects the embeddings into the LLM’s token space. \(\mathbf{b}_p^{(1)} \in \mathbb{R}^{d_{\text{exp}}}\) and \(\mathbf{b}_p^{(2)} \in \mathbb{R}^{d_2}\) are bias terms introduced to enhance expressiveness, ensuring that each transformation incorporates an adaptive shift in feature space.  ReLU activation enables non-linear alignment.

% ~\\
% \noindent $\bullet$ \textbf{Reshaping}  
\paragraph{Reshaping} 

\noindent Since LLMs process text as token sequences, user-item representations must be reshaped to align with the LLM’s tokenization, enabling seamless integration of collaborative signals. To achieve this, the transformed embeddings \((\mathbf{u}_{\text{proj}}, \mathbf{i}_{\text{proj}}) \in \mathbb{R}^{d_2}\) are then reshaped into collaborative tokenization embeddings compatible with the LLM:
\begin{equation}
\begin{aligned}
\mathbf{e}_{x_\text{LLM}} &= \text{Reshape}\left( \mathbf{x}_{\text{proj}} \right) \in \mathbb{R}^{d_2 \times \textbf{proj\_token\_num}},
\end{aligned}
\end{equation} 
\noindent where \(\textbf{proj\_token\_num}\) determines the number of tokens used to represent each user/item.

\begin{figure} % Single column figure
	\includegraphics[width=\linewidth]{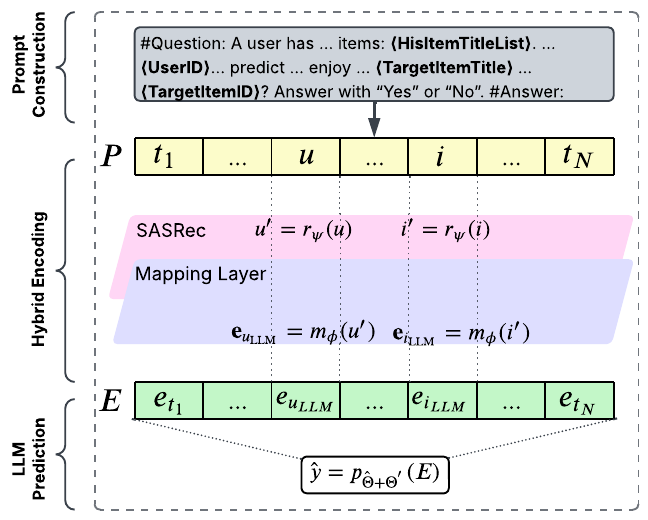}
    \caption{\textbf{Architecture of the LLM Layer.} The LLM layer receives the constructed prompt and aligned collaborative embeddings from SASRec and the mapping layer, and \textbf{generates the final recommendation output.}}
    \label{fig: LLM}
\end{figure}

\subsubsection{LLM}
In addition to collaborative encoding, incorporating an LLM layer into the SASRecLLM architecture is important to leverage the contextual power of LLM and to address the cold start problem. To harness LLMs for SASRecLLM, as shown in Fig. \ref{fig: LLM}, the recommendation data is embedded in LLM-compatible prompts, which are subsequently encoded via a hybrid encoding mechanism. Importantly, a LoRA module is introduced to fine-tune the LLM to perform the recommendation task in an effective and lightweight manner.

% ~\\
% \noindent $\bullet$ \textbf{Prompt Construction} 
\paragraph{Prompt Construction} 

\noindent To convert recommendation data into language prompts for LLM, a fixed prompt template is utilized to construct structured input, drawing inspiration from existing research.

\begin{center}
\fcolorbox{black}{gray!6}{\parbox{0.95\columnwidth}{\noindent \textbf{Prompt Template:} \#Question: A user has given high ratings to the following items: $\langle$\textit{HisItemTitleList}$\rangle$. \uline{Additionally, user preferences are encoded in the feature $\langle$\textit{UserID}$\rangle$}. Using all available information, predict whether the user would enjoy the item titled $\langle$\textit{TargetItemTitle}$\rangle$ \uline{with the feature $\langle$\textit{TargetItemID}}$\rangle$? Answer with ``Yes'' or ``No''. \#Answer:}}
\end{center}
\vspace{-3pt}
\noindent Here, “⟨TargetItemTitle⟩” denotes the title of the item for prediction, while “⟨UserID⟩” and “⟨TargetItemID⟩” incorporate user and item IDs, respectively. To maintain semantic coherence when integrating user/item IDs, they are treated as a feature of users/items within the prompt, as indicated by the underlined content. Moreover, to enhance sequential features and enrich LLMs' semantic understanding, “⟨HisItemTitleList⟩” represents a chronologically ordered list of item titles that a user has interacted with, serving as a textual representation of user preferences. For each recommendation sample, the four fields are populated with corresponding values to generate a structured prompt. Notably, a binary question (“Yes”/”No”) is explicitly appended to build the binary classification task.

% ~\\
% \noindent $\bullet$ \textbf{Hybrid Encoding} 
\paragraph{Hybrid Encoding} 

\noindent A hybrid encoding approach aligns textual and collaborative information. Text is tokenized and embedded using the LLM’s built-in mechanism, while collaborative information is processed using the SASRec model and the mapping layer. For a prompt associated with user $u$ and item $i$, tokenization is first applied using the LLM build-in tokenizer. The tokenized output is denoted as: 
\begin{equation}
P = [t_1, t_2, \dots, t_n, u, t_{n+1}, \dots, i, \dots, t_N],
\end{equation}
where $t_n$ represents a text token, and $u$/$i$ denotes the user/item ID positioned within the ``$\langle$UserID$\rangle$'' / ``$\langle$TargetItemID$\rangle$'' fields.    

The output is then encoded into a sequence of embeddings \( E \):
\begin{equation}\label{eq:hybrid_encoding}
        E = [\bm{e}_{t_1},\dots,\bm{e}_{t_n},\mathbf{e}_{u_\text{LLM}}, \bm{e}_{t_{n+1}}, \dots,\mathbf{e}_{i_\text{LLM}}, \dots, \bm{e}_{t_N}],
\end{equation}
where \( \bm{e}_{t_n} \in \mathbb{R}^{d_{2}} \) represents the token embedding for \( t_n \), obtained via embedding lookup: $\bm{e}_{t_n} = \text{Embedding}_{LLM}(t_n)$. The collaborative tokenization embeddings \( \mathbf{e}_{u_\text{LLM}}, \mathbf{e}_{i_\text{LLM}} \in \mathbb{R}^{d_{2}} \) for user \( u \) and 
% TODO: refine this by mentioning equation
item \( i \) are derived using the SASRec \eqref{eq:sasrec}, followed by reshaping via the mapping layer \eqref{eq:mapping_layer}.

\paragraph{LLM Prediction} 
\noindent Once the input prompt is converted into an embedding sequence \(E\), derived from Equation \eqref{eq:hybrid_encoding}, the LLM uses it to generate the final results. However, purely relying on LLM lacks recommendation-specific training, a LoRA module is introduced to enhance predictive capabilities. The prediction is defined as:  
\begin{equation}\label{eq:llm_predict}
\hat{y} = p_{\hat{\Theta}+\Theta^{'}}(E),
\end{equation}
where \(\hat{\Theta}\) represents the fixed parameters of the LLM \(p(\cdot)\), and \(\Theta^{'}\) denotes the learnable LoRA parameters for the recommendation task, \(\hat{y}\) corresponds to the probability of the label being \(1\), indicating the likelihood of the LLM answering "Yes."

\begin{figure*}[t] % Two column figure (notice the starred environment)
	\includegraphics[width=\linewidth]{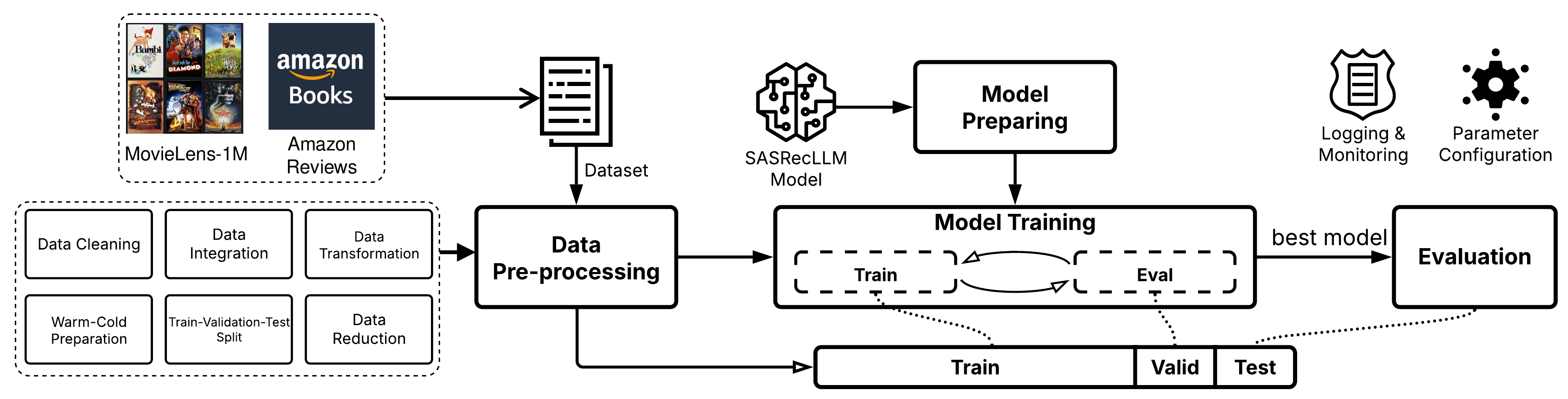}
	\caption{Overview of the SASRecLLM implementation workflow.}
    \label{fig:impl}
\end{figure*}

\subsection{Training strategies} \label{sec: training_strategies}
Training the model parameters requires a structured approach. A naive method would be to train all components simultaneously. However, due to the strong dependence on structured collaborative representations, the jointly optimizing of all layers from scratch can lead to suboptimal learning, particularly in cold start scenarios where the system lack sufficient exposure to recommendation-specific data.  Additionally, computational efficiency is a key concern, as modern LLM-based systems require resource-efficient tuning strategies to remain practical for large-scale development. Moreover, as a multi-layer hybrid architecture, SASRecLLM requires a modular training approach to improve adaptability and avoid optimization conflicts between different components \cite{chatgpt}. To address these challenges, three complementary training strategies are proposed.

\subsubsection{Dual-Stage Training} \label{sec: dual_stage_train}
Dual-Stage Training involves two distinct phases: pre-training and full-system fine-tuning. In the pre-training phase, the SASRec and LLM components are trained independently to ensure each learns effectively before integration. In the subsequent phase, these pre-trained components are combined and fine-tuned jointly to optimize overall system performance.

Specifically, SASRec is first trained on numerical interaction data in isolation. This step enables it to generate high-quality collaborative representations, minimizing the risk of unrefined embeddings and unstable gradients during joint training. The training process follows a transformer-based architecture with a standard train-eval loop, as described in Section \ref{sec:implementation}. Simultaneously, the LLM undergoes fine-tuning using textual-only prompts to warm up for recommendation tasks. During this stage, user-item IDs are excluded from the prompt, ensuring that the LLM can specializes in text-based recommendation learning. This phase leverages the LLM’s natural language understanding capabilities, allowing it to better contextualize inputs before being exposed to collaborative signals. 

After finishing the pre-trained phase, an additional fine-tuning stage further aligns it with LLM tokenization, enhancing adaptability to the hybrid architecture and ultimately improving recommendation accuracy. During this stage, the mapping layer is also optimized to perform alignment more effectively. Overall, this method training method improves both learning efficiency and accuracy during training.

\subsubsection{Hierarchical Freezing}
This method enables independent fine-tuning of each component in SASRecLLM by freezing the parameters of others during training, thereby enhancing modularity and reusability. It supports the Dual-Stage Training strategy by selectively freezing or unfreezing modules at strategic stages, which prevents interference between optimization objectives and ensures more stable and efficient parameter updates.

\subsubsection{Plug-and-Play Tuning}
The goal of tuning is to optimize model parameters to minimize the loss. Formally, for a data point $(u, i, y)$ in the historical interaction dataset $\mathcal{D}$, where $u$ and $i$ correspond to a user and an item, respectively, with $y\in\{1,0\}$ indicating the interaction label ("Yes" or "No"), this corresponds to solving the following optimization problem:
\begin{equation}
\min_{\Omega} \sum_{(u,i,y)\in \mathcal{D}} \ell(\widehat{y},y), 
\quad \Omega=\{\Theta^{'},\phi,\psi\},
\end{equation}
where $\widehat{y}$ corresponds to the model’s predicted probability, derived from Equation \eqref{eq:llm_predict}, and $\Omega$ represents the set of trainable parameters: $\Theta^{'}$ for the LoRA module, $\phi$ for the mapping layer, and $\psi$ for the SASRec encoder. The loss function $\ell$ is defined as binary cross-entropy (BCE), commonly used for binary classification:
\begin{equation}
\ell(y, \widehat{y}) = - \left[ y \log \widehat{y} + (1 - y) \log (1 - \widehat{y}) \right].
\end{equation}

Plug and Play (PnP) tuning allows the SASRecLLM framework to flexibly load and fine tune individual components independently while preserving the overall system architecture. In contrast to  end-to-end tuning, which updates all parameters at once, PnP tuning enables the isolation of each module during training. By isolating and fine-tuning individual components, this method minimizes computational overhead compared to full-model retraining, making it resource-efficient for large-scale systems. 

\begin{figure*}[t] % Two column figure (notice the starred environment)
	\includegraphics[width=\linewidth]{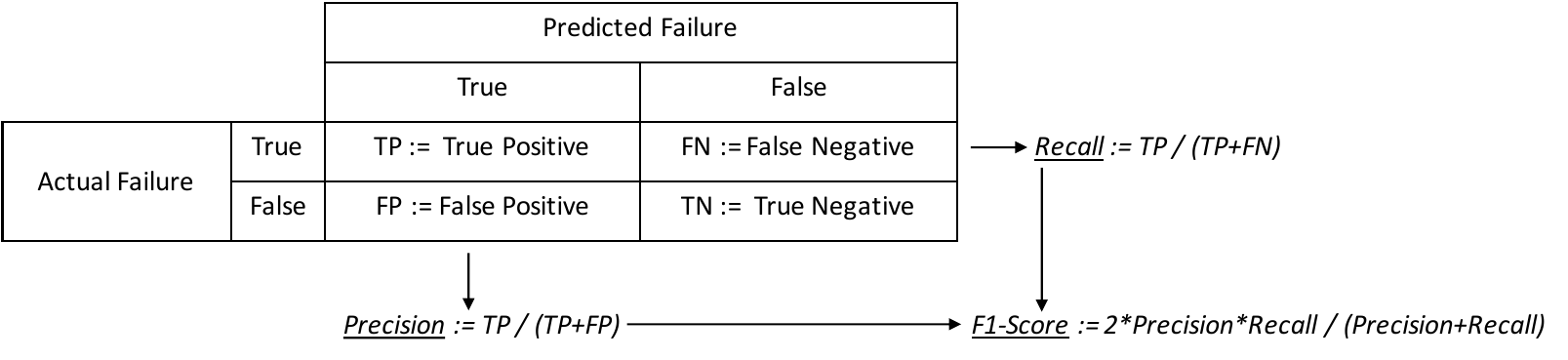}
    \caption{Illustration of Confusion Matrix-Based Metrics \cite{spiegel2018cost}.  
    \( TP \): Correctly predicted failures.  
    \( TN \): Correctly predicted non-failures.  
    \( FP \): Incorrectly predicted failures.  
    \( FN \): Missed failures.}
    \label{fig: metrics}
\end{figure*}

\section{Implementation} \label{sec:implementation}

SASRecLLM is implemented in Python and hosted on Kaggle Notebook, utilizing its free T4 GPU quota (30 hours per week). This setup reduces platform dependency while enhancing centralized management and efficient distribution. All machine learning components are implemented using PyTorch 2.0, whereas LLM-related components are sourced from Hugging Face. Fig. \ref{fig:impl} illustrates the overall workflow.

% This table will span across both columns at the top of a page.
% \begin{table*}[b] 
% \centering
% \renewcommand\arraystretch{1.2}
% \caption{Statistics of the Preprocessed Datasets. Values indicate the number of samples in each subset.}
% \label{tab: dataStatics}
% % CORRECTED: Now has 'l' + 7 'c's for a total of 8 columns.
% \begin{tabular*}{\textwidth}{@{\extracolsep{\fill}}lccccccc} 
% \toprule
% \textbf{Sub-Dataset} & \textbf{\#Train} & \textbf{\#Valid} & \textbf{\#Test} & \textbf{\#User} & \textbf{\#Item} & \textbf{\#Cold} & \textbf{\#Warm} \\
% \midrule
% MovieLens-1M & 33,891 & 10,401 & 7,331 & 839 & 3,256 & 26 & 2,085 \\
% Amazon-Book & 11,174 & 2,394 & 2,395 & 4,000 & 4,001 & 310 & 2,085 \\
% \bottomrule
% \end{tabular*}
% \end{table*}

\subsection{Data Engineering}
\subsubsection{Dataset}
~\\
$\bullet$ \textbf{MovieLens-1M \cite{movielens1m_kaggle}} A widely used dataset containing movie details, user ratings, and tags, supporting the development and evaluation of personalized recommendation algorithms.

$\bullet$ \textbf{Amazon Book Reviews  \cite{ni2019amazon}} This study utilizes the book subset of the Amazon Product Review dataset. This subset contains approximately 3 million book reviews from 212,404 unique books and their respective reviewers, collected between 1996 and 2018. Reviews are rated on a 1 to 5 scale, reflecting user preferences.

\subsubsection{Data Pre-processing}
To prepare the dataset for this study, several preprocessing steps are applied. 
% Table~\ref{tab: dataStatics} presents the statistical summary of the processed dataset.

\noindent $\bullet$  \textbf{Data Cleaning}  

\noindent This step ensures data quality by correcting inconsistencies and removing incomplete records to prevent training errors, reducing dataset size while retaining meaningful training samples.

\noindent $\bullet$ \textbf{Data Integration}

\noindent  Multiple sources of information (ratings, item metadata, and user attributes) are merged into a single structured dataset, enabling more comprehensive analysis and improving model performance.

\noindent $\bullet$ \textbf{Data Transformation}  

\noindent To improve interpretability and model compatibility, raw data undergoes standardization and reformatting. This process includes normalizing numerical features, mapping categorical variables, and ensuring consistent data representation. User ratings ranging from 1 to 5 are converted into binary labels using a threshold of 4 to facilitate binary classification. Ratings greater than or equal to 4 are labeled as “like” (\(y = 1\)), while those below 4 are labeled as “dislike” (\(y = 0\)).

\noindent $\bullet$ \textbf{Data Reduction} \label{sec:data_reduction}

\noindent To manage computational constraints, data reduction is applied to retain only the most relevant information while minimizing dataset size. For the Amazon Book dataset, the original size exceeds current hardware memory limits, so records with user and item IDs greater than 4,000 are filtered out to reduce computational overhead.

\noindent $\bullet$ \textbf{Train-Validation-Test Split}  

\noindent The datasets are partitioned into train, validation, and test subsets in an 8:1:1 ratio. The training set learns model parameters, the validation set fine-tunes hyperparameters, and the test set assesses final performance. This partitioning ensures that future interactions do not appear in the training data, preventing potential information leakage.

\noindent $\bullet$ \textbf{Warm-Cold Preparation}  \label{sec:cold_warm} 

\noindent To evaluate SASRecLLM’s performance under both cold-start and warm-start scenarios, two additional test sets are constructed. The \textbf{warm} set includes user-item pairs where the interaction count exceeds a predefined threshold of 3, ensuring that users have sufficient historical interactions. In contrast, the \textbf{cold} set consists of samples where user-item interactions are absent or minimal, simulating a cold-start scenario where the model must make predictions with little or no prior user behavior data.

\subsection{Model Preparation}
The implementation of \textbf{SASRec} adopts a Transformer-based SRS that uses self-attention to model user-item interactions. The model consists of two Transformer blocks (\(b=2\)), each with four attention heads per layer. Both item embeddings and positional embeddings are initialized with a dimensionality of 64. The Point-Wise FFN consists of two Conv1D layers with ReLU activation. Layer Normalization is used to stabilize training, and a dropout rate of 0.2 is applied to prevent overfitting. Training is conducted with a batch size of 1028, and the maximum sequence length \(n=25\) is set to approximate the mean number of interactions per user. The optimizer used is Adam, with learning rate scheduling with \(1e-2\).

The \textbf{Mapping Layer}  
is implemented as a two-layer MLP, this module expands embeddings to a higher-dimensional space, then projects them to the LLM’s embedding size. Reshaping is performed using PyTorch’s reshape function.

The backbone of \textbf{LLM} is LLaMA, and a LoRA module is introduced for fine-tuning. Among various LLMs, SASRecLLM selects LLaMA for its open-source nature, efficiency, and fine-tunable attributes. Among the available LLaMA models, SASRecLLM chooses TinyLlama-1.1B\cite{zhang2024tinyllama} due to limited computational resources. With just 1.1B parameters, TinyLlama is a compact model designed for applications requiring a low computational and memory footprint. For the LoRA model, SASRecLLM follows the same configuration as described in the TALLRec paper \cite{bao2023tallrec}.

\subsection{Strategy Implementation}
The checkpoint feature serves as the foundation for implementing all strategies, which capture a model's internal state, including weights, biases, and other parameters, at a particular point in the training process. A frequency parameter of 8 is explicitly configured for saving checkpoints, meaning a checkpoint is saved every 8 epochs. The best-performing checkpoint across all epochs is also retained. 

\textbf{Dual-Stage Training} is implemented by first training the SASRec model independently, saving the best-performing model, and then loading it into SASRecLLM. \textbf{Hierarchical Freezing} is implemented by disabling gradient updates for the frozen layers, ensuring that their parameters remain unchanged during backpropagation. The checkpoint feature and Hierarchical Freezing together enable the implementation of \textbf{PnP Tuning}. When tuning LoRA with textual-only data, SASRecLLM freezes all other layers. After tuning the LoRA layer, SASRecLLM saves a checkpoint, resumes the system, reloads the LoRA checkpoint, freezes the pre-trained LoRA layer, and fine-tunes the remaining layers.

\subsection{Model Training}
The training process of SASRecLLM follows a \textbf{Train-Eval Pattern}, implemented through a modular \texttt{Runner} class for structured execution. Each epoch alternates between \textbf{training} and \textbf{evaluation}, ensuring model stability and convergence.

During \textbf{training}, SASRecLLM processes input data through prompt construction and hybrid encoding before computing logits, the raw, unnormalized output from the final layer of the model. The \texttt{Runner} class automates batch processing via distributed data loaders with a batch size of 16 while optimizing parameters using AdamW, which applies weight decay to prevent overfitting and ensures stable weight updates. Gradient updates are performed with mixed precision using \texttt{GradScaler}, enhancing memory efficiency and preventing gradient underflow. To further stabilize training and improve generalization, a learning rate scheduler is used, initially applying a warm-up phase before smoothly decaying the learning rate using a cosine schedule. BCE loss is applied, and gradients are updated through backpropagation.

During \textbf{evaluation}, the model is assessed on the validation set using predefined metrics to log performance statistics for subsequent training analysis, more details will be discussed in Section \ref{sec:evaluation}). The \texttt{Runner} class integrates early stopping by monitoring validation metrics and stopping training when no further improvement is observed, thus reducing the risk of overfitting. Notably, the Runner class conducts the final evaluation separately on an independent test set after training, ensuring an unbiased assessment of generalization performance.

Upon completion, the best-performing model is saved as a checkpoint. Training is conducted for a maximum of 300 epochs. In the 1×T4 setup, SASRecLLM requires 9 hours to train on a single dataset (either MovieLens or Amazon Book).

\section{Experiments} \label{sec:evaluation}

This section evaluates SASRecLLM on the previous two benchmark datasets by addressing the following research questions to validate its effectiveness and superiority:

\noindent \textbf{RQ1:} Can the proposed SASRecLLM outperform baseline methods?

\noindent \textbf{RQ2:} What are the effects of different components within SASRecLLM?

\noindent \textbf{RQ3:} How does SASRecLLM perform in cold-start and warm-start scenarios?

\noindent \textbf{RQ4:} What insights can be drawn from the training process of SASRecLLM?

\subsection{Baselines}
To comprehensively evaluate SASRecLLM's performance and compare it with existing CRMs and LLM4Rec, several baselines are proposed and implemented, they are categorized into three groups:

\subsubsection{CF-Based Baselines}
~\\
$\bullet$ \textbf{MF} 
An important CF method utilizes user-item interaction data to uncover preference patterns \cite{wang2024trustworthy}. It decomposes the large user-item interaction matrix into two smaller matrices, revealing hidden information about user preferences and item properties \cite{bokde2015matrix}.

\noindent $\bullet$ \textbf{Neural Collaborative Filtering}  
This baseline extends MF using deep learning and employs an MLP to model user-item interactions. It captures non-linear relationships between users and items and is more expressive than simple dot-product-based MF \cite{he2017neural}.

\subsubsection{SRSs-Based Baselines}
~\\
\noindent $\bullet$ \textbf{RNN \cite{hidasi2016session}}  
This method is designed to process sequences, making it ideal for tasks involving evolving user behavior or preferences. Rather than relying on long-term user histories, it excels in session-based scenarios where recommendations are generated from short-term interactions within a single session. Additionally, RNN can account for varying time intervals between events, adapting recommendations as user preferences change over time. 

\noindent $\bullet$ \textbf{MC} 
This baseline models sequential item interactions based on transition probabilities. A state represents a user's current interacted items, and the transition probability defines the likelihood of moving from one item to another based on past interactions. Given a user's history, this baseline predicts the most probable next item for recommendation \cite{shani2005mdp}.

\noindent $\bullet$ \textbf{SASRec}  
Although integrated into SASRecLLM, SASRec serves as a strong baseline to compare standalone SRS with the proposed framework. The detailed comparison is provided in the Ablation Study Section \ref{sec: abl}.

\subsubsection{LLM4Rec-Based Baselines}
~\\
\noindent $\bullet$ \textbf{ICL} 
This baseline leverages LLM’s NLU capability to generate recommendations by directly querying the original model with prompts \cite{brown2020language}.

\noindent $\bullet$ \textbf{TALLRec}  
This baseline aligns LLMs with recommendation tasks via LoRA fine-tuning \cite{bao2023tallrec}. It represents a simplified variant of the proposed framework, as it fine-tunes the LLM without incorporating sequential collaborative information. This baseline is also evaluated as part of the ablation study.

To ensure fairness across different baseline groups, all models are implemented under the same experimental settings with identical datasets, as described in Section \ref{sec:implementation}. Specifically, CF-based baselines are trained and evaluated solely on user-item interaction data, while SRSs-based methods incorporate users’ historical interaction sequences. All LLM4Rec models share the same TinyLlama-1.1B backbone, employ the same LoRA fine-tuning configuration, and use a consistent prompt template. Unlike deterministic baselines, LLM4Rec models exhibit stochastic inference behavior due to their internal architecture and computation patterns. To ensure robust and reliable evaluation, those models are assessed multiple times, and the average score is reported. In contrast, CF-based and SRSs-based baselines require only a single evaluation run due to their deterministic nature.

\begin{figure*}[h] % Two column figure (notice the starred environment)
    \includegraphics[width=\linewidth]{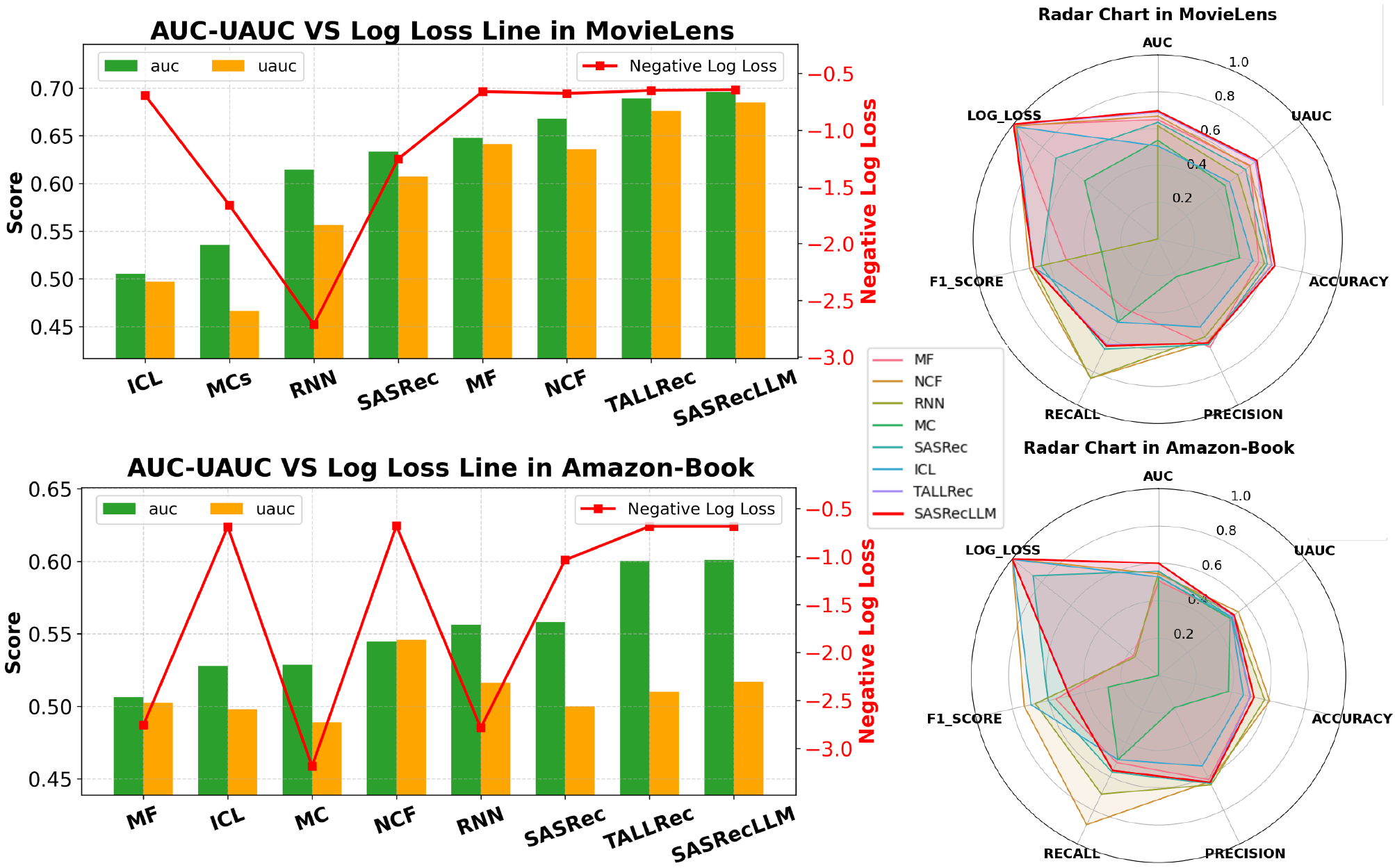}
    \caption{
        Comprehensive performance comparison of SASRecLLM and baseline models on the MovieLens (top) and Amazon Book (bottom) datasets across multiple evaluation metrics. Each row presents:
        (1) AUC, UAUC (bars), and Negative Log Loss (line) are used to assess recommendation accuracy. From left to right, the graph is sorted by AUC.
        (2) a radar chart visualizing model consistency across all metrics.
    }
    \label{fig: baseline_com}
    
\end{figure*}

\subsection{Metrics}
To evaluate SASRecLLM against baselines, three widely adopted metrics in RSs are used as the primary metrics: AUC and UAUC for ranking effectiveness, and Log Loss for probabilistic prediction quality. AUC and UAUC capture the model’s capacity to distinguish between positive and negative interactions, while Log Loss penalizes confident but incorrect predictions, offering a finer-grained view of predictive reliability. In this context, even small improvements such as a 0.001 increase in AUC/UAC or a 0.001 decrease in Log Loss are considered practically meaningful. Additionally, confusion matrix-based metrics (e.g., accuracy, precision, recall, and F1-score) are included as secondary metrics to provide a more comprehensive and interpretable performance analysis. Time-related metrics are excluded, as they are highly sensitive to hardware configurations and not directly indicative of model quality.

\subsubsection{AUC \& UAUC}
~\\
\noindent $\bullet$ \textbf{AUC}
The Area Under the ROC Curve (AUC) measures the model’s ability to predict user preferences by quantifying the trade-off between true positive and false positive rates. AUC evaluates how well the model ranks relevant items above irrelevant ones, with a higher score (closer to 1) indicating better recommendation performance, while 0.5 represents random guessing \cite{bowers2019receiver}. AUC is computed as: 
\begin{equation}
\text{AUC} = \frac{1}{|\mathcal{P}| |\mathcal{N}|} \sum_{p \in \mathcal{P}} \sum_{n \in \mathcal{N}} \mathbf{f_\text{auc}}(s_p > s_n),
\end{equation}
where \( \mathcal{P} \) and \( \mathcal{N} \) represent sets of positive (like) and negative (dislike) samples, respectively. \( s_p \) and \( s_n \) denote their predicted scores. The function \( \mathbf{f_\text{auc}}(s_p > s_n) \) returns 1 if the model assigns a higher score to the positive sample than the negative one, indicating a correct ranking; otherwise, it returns 0.

\noindent $\bullet$ \textbf{UAUC}
User-wise AUC (UAUC) extends AUC by computing the ranking quality per user and then averaging across users, making it more robust for recommendation tasks with varying user behavior. Unlike standard AUC, which considers overall pairwise ranking, UAUC evaluates personalized ranking performance at the user level \cite{jimenez2022uniform}. Formally,
\begin{equation}
\text{UAUC} = \frac{1}{|\mathcal{U}|} \sum_{u \in \mathcal{U}} \frac{1}{|\mathcal{P}_u| |\mathcal{N}_u|} \sum_{p \in \mathcal{P}_u} \sum_{n \in \mathcal{N}_u} \mathbf{f_\text{uauc}}(s_p > s_n),
\end{equation}
where \( \mathcal{U} \) is the set of users, and \( \mathcal{P}_u \), \( \mathcal{N}_u \) are the positive and negative interactions for user \( u \). The ranking score \( s_p \) and \( s_n \) are compared within each user's scope. A higher UAUC indicates better-personalized performance, making it a crucial metric for SASRecLLM.

\subsubsection{Confusion Matrix-Based Evaluation Metrics}
~\\
\noindent $\bullet$ \textbf{Precision}
This metric measures the proportion of correctly predicted relevant items, reflecting recommendation accuracy. A higher value indicates fewer false positives and greater reliability.

\noindent $\bullet$ \textbf{Recall}  
It measures the model's ability to identify relevant items. Higher recall reduces missed recommendations, crucial when maximizing relevant item retrieval.

\noindent $\bullet$ \textbf{F1-Score}  
balances precision and recall using their harmonic mean. A higher F1-score reflects better accuracy, crucial when both missing relevant items and incorrect recommendations matter.

\noindent $\bullet$ \textbf{Accuracy}  
measures the proportion of correctly classified predictions.

\begin{table*}[ht]
\centering
\small
\caption{Overall Performance Comparison on the \textbf{MovieLens} Dataset. All metrics are reported to three decimal places, as even a 0.001 change is considered practically meaningful. \textbf{“Rel. Imp.”} denotes the relative improvement of SASRecLLM over each baseline, averaged across \textbf{AUC} and \textbf{UAUC}.}
\label{tab:metrics_movielens_grouped}
\vspace{-5pt}
%--- START CHANGES ---
\resizebox{\textwidth}{!}{
%--- END CHANGES ---
\begin{tabular}{@{}llcccccccc@{}}
\toprule
& & & \multicolumn{4}{c}{\textbf{Confusion Matrix-Based}} & \multicolumn{3}{c}{\textbf{AUC-based Metrics}} \\
\cmidrule(lr){4-7} \cmidrule(lr){8-10}
\textbf{Category} & \textbf{Model} & \textbf{Log Loss} & \textbf{Precision} & \textbf{Recall} & \textbf{F1-Score} & \textbf{Accuracy} & \textbf{AUC} & \textbf{UAUC} & \textbf{Rel. Imp. (\%)} \\
\midrule
\multirow{2}{*}{CF-based}
& MF & 0.658 & 0.652 & 0.417 & 0.509 & 0.561 & 0.648 & 0.641 & 7.14\% \\
& NCF & 0.675 & 0.620 & 0.839 & 0.713 & 0.632 & 0.668 & 0.636 & 5.93\% \\
\midrule
\multirow{3}{*}{SRSs-based}
& RNN & 2.713 & 0.589 & 0.838 & 0.692 & 0.593 & 0.614 & 0.557 & 17.96\% \\
& MCs & 1.661 & 0.227 & 0.500 & 0.313 & 0.455 & 0.536 & 0.467 & 37.84\% \\
& SASRec & 1.252 & 0.636 & 0.663 & 0.649 & 0.609 & 0.634 & 0.607 & 11.31\% \\
\midrule
\multirow{3}{*}{LLM4Rec-based}
& ICL & 0.692 ± 0.006 & 0.529 ± 0.048 & 0.500 ± 0.000 & 0.691 ± 0.041 & 0.528 ± 0.047 & 0.505 ± 0.010 & 0.497 ± 0.014 & 37.76\% \\
& TALLRec & 0.649 ± 0.063 & 0.628 ± 0.116 & 0.636 ± 0.048 & 0.688 ± 0.078 & 0.635 ± 0.064 & 0.689 ± 0.052 & 0.676 ± 0.059 & 1.17\% \\
\rowcolor{green!15}
& \textbf{SASRecLLM} & \textbf{0.642 ± 0.031} & \textbf{0.626 ± 0.079} & \textbf{0.644 ± 0.035} & \textbf{0.691 ± 0.053} & \textbf{0.650 ± 0.029} & \textbf{0.696 ± 0.037} & \textbf{0.685 ± 0.031} & \textbf{---} \\
\bottomrule
\end{tabular}
%--- Add this closing brace ---
}
\end{table*}

\begin{table*}[t] % Use table* to span two columns. [t] places it at the top of the page.
\centering
\small
\caption{Overall Performance Comparison on the \textbf{Amazon Book} Dataset}
\label{tab:metrics_amazon_grouped}
\vspace{-5pt}
% We can still use resizebox to ensure it fits perfectly within the text width of the page
\resizebox{\textwidth}{!}{
% NOTE: Vertical lines have been removed and \hline replaced with booktabs rules for better aesthetics
\begin{tabular}{llcccccccc}
\toprule
\textbf{Category} & \textbf{Model} & \textbf{Log Loss} & \multicolumn{4}{c}{\textbf{Confusion Matrix-Based Evaluation Metrics}} & \multicolumn{3}{c}{\textbf{AUC-based Metrics}} \\
\cmidrule(lr){4-7} \cmidrule(lr){8-10}
& & & \textbf{Precision} & \textbf{Recall} & \textbf{F1-Score} & \textbf{Accuracy} & \textbf{AUC} & \textbf{UAUC} & \textbf{Rel. Imp. (\%)} \\
\midrule

\multirow{2}{*}{CF-based} 
& MF & 2.756 & 0.618 & 0.517 & 0.563 & 0.504 & 0.506 & 0.503 & 10.80\% \\
& NCF & 0.680 & 0.630 & 0.886 & 0.736 & 0.608 & 0.545 & 0.546 & 2.49\% \\
\midrule

\multirow{3}{*}{SRSs-based} 
& RNN & 2.786 & 0.648 & 0.704 & 0.675 & 0.581 & 0.556 & 0.516 & 4.23\% \\
& MC & 3.186 & 0.191 & 0.500 & 0.276 & 0.382 & 0.529 & 0.489 & 9.85\% \\
& SASRec & 1.036 & 0.642 & 0.574 & 0.606 & 0.503 & 0.558 & 0.500 & 5.65\% \\
\midrule

\multirow{3}{*}{LLM4Rec-based} 
& ICL & 0.690 ± 0.003 & 0.538 ± 0.022 & 0.500 ± 0.000 & 0.699 ± 0.018 & 0.464 ± 0.020 & 0.528 ± 0.024 & 0.498 ± 0.026 & 8.96\% \\
& TALLRec & 0.649 ± 0.063 & 0.628 ± 0.116 & 0.636 ± 0.048 & 0.688 ± 0.078 & 0.635 ± 0.064 & 0.689 ± 0.052 & 0.676 ± 0.059 & 1.17\% \\
\rowcolor{green!15}
& \textbf{SASRecLLM} & \textbf{0.685 ± 0.013} & \textbf{0.634 ± 0.044} & \textbf{0.564 ± 0.027} & \textbf{0.489 ± 0.040} & \textbf{0.523 ± 0.032} & \textbf{0.601 ± 0.041} & \textbf{0.517 ± 0.037} & \textbf{---} \\
\bottomrule
\end{tabular}
}
\end{table*}

\subsection{Results} \label{sec: results}

\subsubsection{Performance Comparison (RQ1)} \label{sec: overall_com}

The overall performance on two datasets is summarized in Table \ref{tab:metrics_amazon_grouped} and \ref{tab:metrics_movielens_grouped}, leading to the following observations: 

SASRecLLM demonstrates strong performance across two datasets on the primary metrics, highlighting the effectiveness of the proposed framework. Regarding the secondary metrics, it exhibits strong and consistent performance, providing further evidence of the robustness and reliability of the framework.

Compared to the MovieLens dataset, all baseline models perform slightly worse on the Amazon Book dataset. This can be attributed to the data reduction step, as described in Section \ref{sec:data_reduction}, user IDs greater than 4000 were filtered out to reduce computational overhead. While this step is necessary under the current resource constraints, it may have inadvertently impacted the quality and diversity of user–item interactions in the book dataset.

\begin{figure}[h] % Single column figure
	\includegraphics[width=\linewidth]{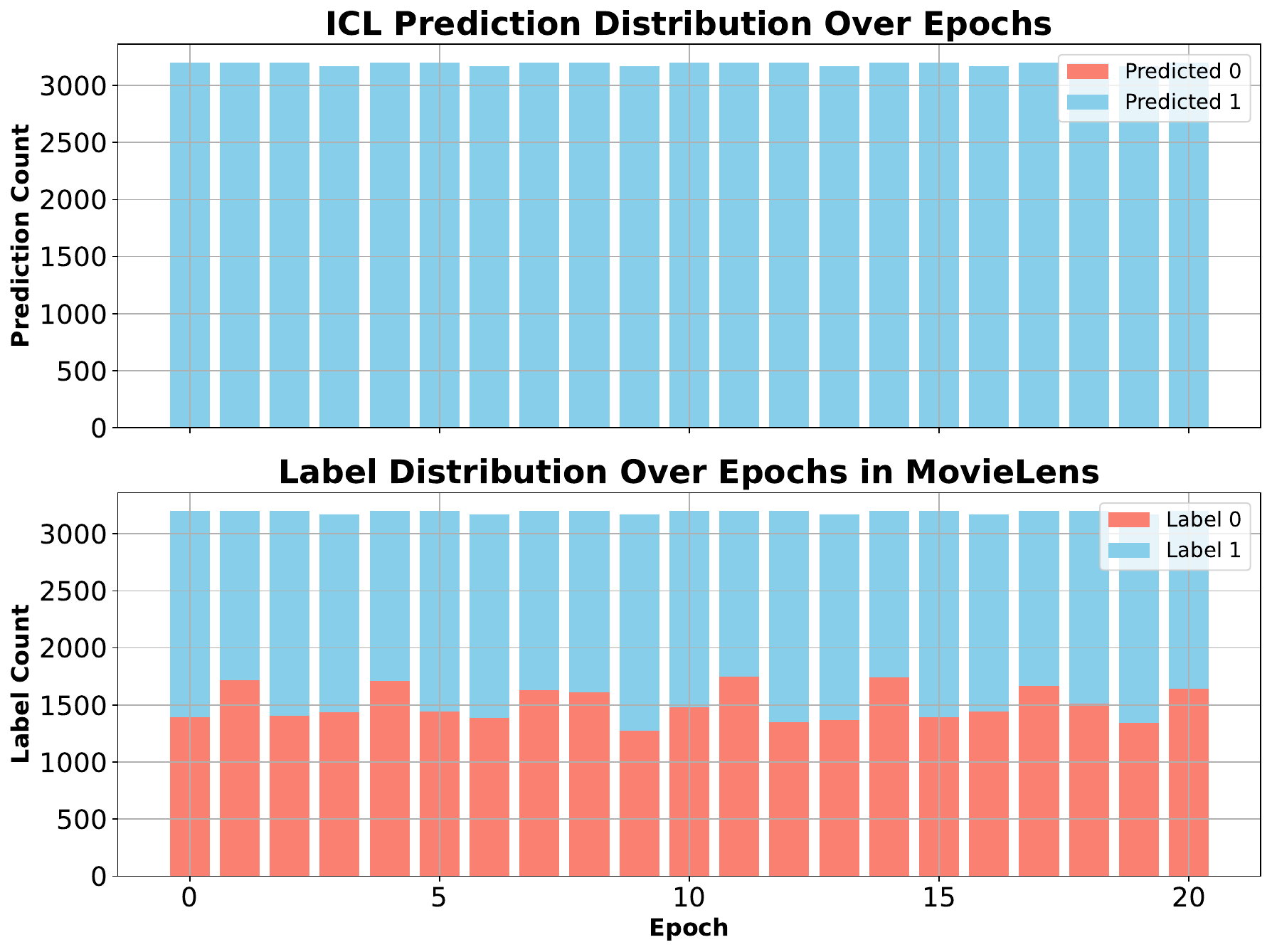}
    \caption{The top chart shows that ICL consistently predicts only class 1 (light blue), while the bottom chart displays the actual label distribution in MovieLens, where both classes (0 and 1) are relatively balanced and vary over time.}
    \label{fig: ICL_dis}
\end{figure}

The ICL baseline exhibits limited performance under the current setting, partly due to the use of TinyLlama-1.1B. Its relatively small parameter size may lack the capacity to effectively model complex recommendation patterns, thus impacting LLM performance. However, to control variance and ensure fairness, it was necessary to select it as the LLM backbone for the ICL baseline. Moreover, as shown in Fig. \ref{fig: ICL_dis}, when evaluate the ICL baseline in multiple runs, the recall rate is always 0.5000 in the balanced dataset. This behavior suggests that, the model only catches the true positives regardless of the input, which confirms the research gap identified in Section \ref{sec: llm4rec}, thus proving the importance of fine-tuning LLMs to teach them to do recommendation tasks. By contrast, the other two LLM4Rec methods effectively address these shortcomings and demonstrate superior predictive performance.

The CF-based baselines perform competitively under the current experimental binary classification setup, whereas SRS-based models are typically more effective when evaluated using sequence-aware or ranking-oriented metrics (e.g., Hit@K, Mean Reciprocal Rank). This evaluation context, therefore, aligns more closely with the strengths of CF methods. Nonetheless, SRS-based techniques excel at modeling sequential user behavior and capturing dynamic preferences, as discussed in Section~\ref{sec: RSs}. To enhance flexibility, future work could allow configurable CRM encoders and explore broader setups. 

Among the SRS-based baselines, SASRec demonstrates competitive overall performance, attributed to its self-attention mechanism for adaptively modeling user–item interactions. Based on this strength, the proposed framework adopts SASRec as the collaborative encoder backbone.

\subsubsection{Ablation Study (RQ2)} \label{sec: abl}

This section presents an ablation study to evaluate the contributions of SASRecLLM’s key components: the standalone SASRec module and the fine-tuned LLM without the CRM encoder (referred to as TALLRec). The variation without the mapping layer is not evaluated, as the mapping layer is essential for aligning the embedding spaces of SASRec and the LLM, making its removal infeasible.

% \begin{figure*} % Two column figure (notice the starred environment)
% 	\includegraphics[width=\linewidth]{Figures/ab_study_col.pdf}
%     \caption{Relative improvement line chart for ablation study using ICL as the benchmark.}
%     \label{fig: ablation-study}
% \end{figure*}

\begin{figure}[h] % Single column figure
	\includegraphics[width=\linewidth]{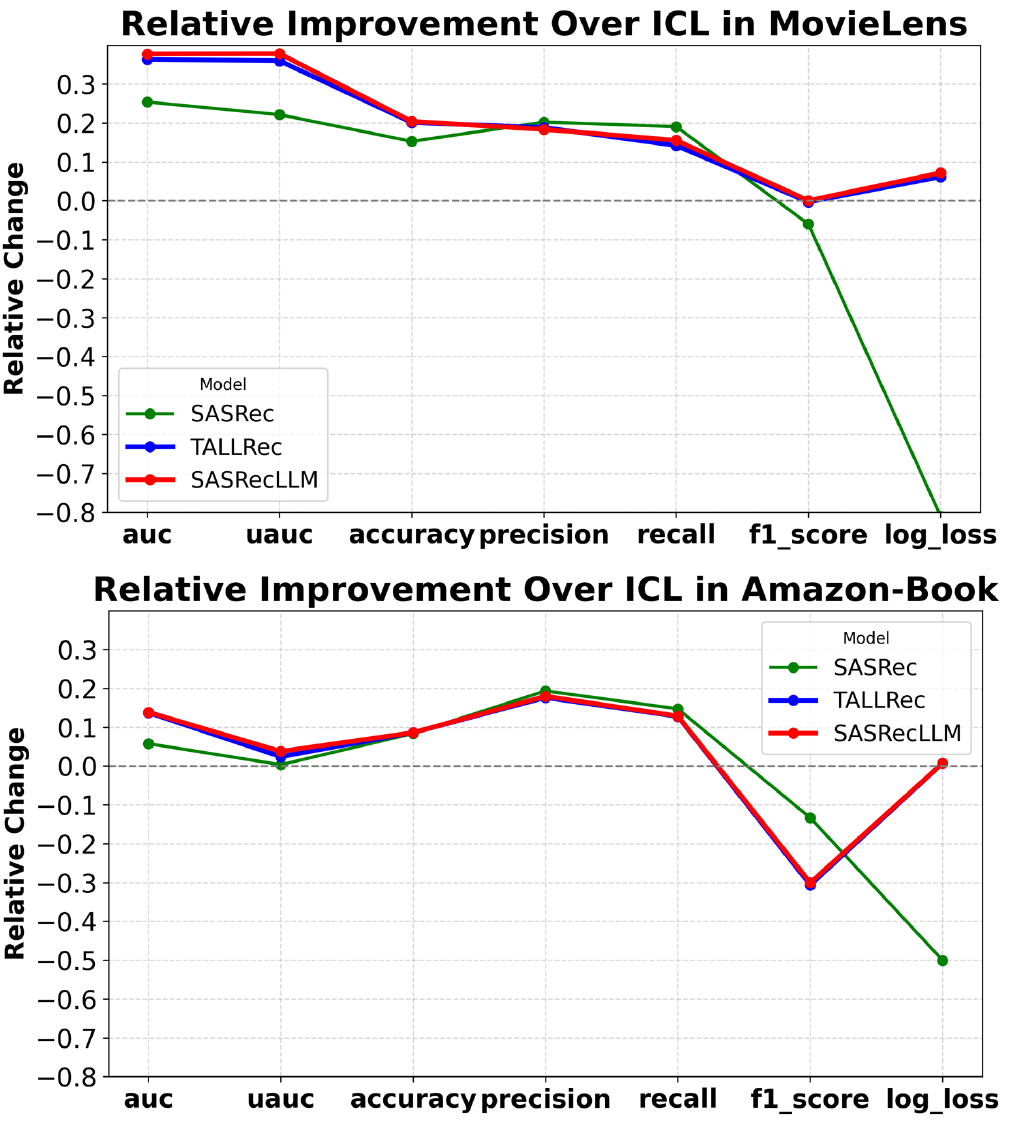}
    \caption{Relative improvement line chart for ablation study using ICL as the benchmark.}
    \label{fig: ablation-study}
\end{figure}

Fig. \ref{fig: ablation-study} summarizes the ablation results across both datasets. As shown by the dashed line in the figure, ICL is chosen as the benchmark for improved visualization and its inherent suitability as the primary LLM4Rec method. SASRecLLM consistently achieves the best performance across all evaluation metrics, validating the effectiveness of the integrated architecture. Notably, the largest performance gain is observed between the standalone SASRec model and SASRecLLM, underscoring the power of LLM4Rec. In comparison, the performance improvement from TALLRec to SASRecLLM is relatively modest. This suggests that fine-tuning alone can yield substantial benefits, and further investigation is needed to better understand the benefit of integrating collaborative signals into SASRecLLM.

\subsubsection{Cold and Warm Performance (RQ3)}

\begin{figure}[h] % Single column figure
	\includegraphics[width=\linewidth]{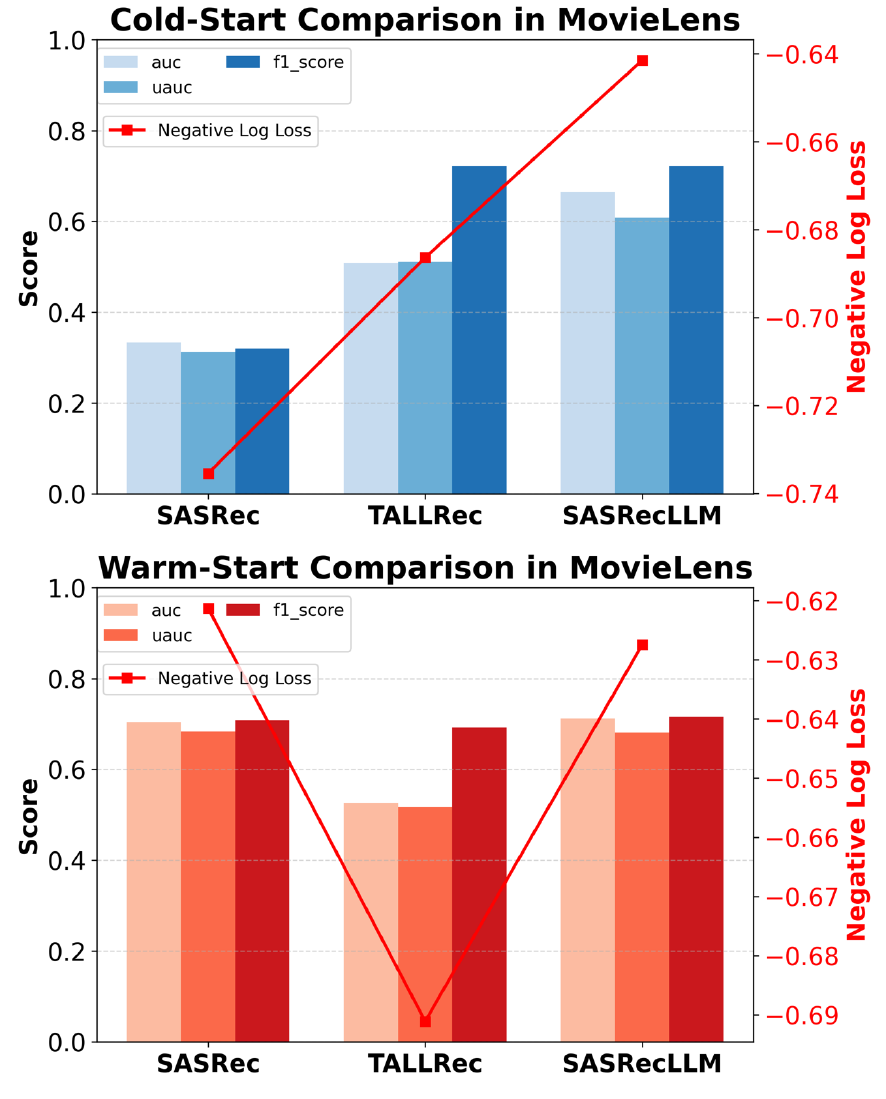}
    \caption{Cold and Warm Comparison in MovieLens.}
     \label{fig: cold_warm}
\end{figure}

The findings from RQ2 motivate further investigation into the complementary benefits of integrating fine-tuned LLMs with collaborative signals. Prior research has shown that LLM4Rec is particularly effective in cold-start scenarios because of LLM’s generalization capabilities and world knowledge \cite{bao2023tallrec}, while CRMs are advantageous in warm settings where user-item interactions are rich \cite{aggarwal2016collaborative}. SASRecLLM aims to bridge these strengths, incorporating collaborative information into LLM4Rec to enhance performance across both scenarios. To evaluate this objective, the performance of SASRec, TALLRec, and SASRecLLM is compared under cold and warm conditions using the pre-processed dataset described in Section~\ref{sec:cold_warm}. Given similar trends and higher data quality, MovieLens is used as the focus. In addition to primary metrics, the F1 Score is included from the secondary metrics set to ensure a fair and comprehensive comparison. As shown in Fig.~\ref{fig: cold_warm}, several key observations emerge:

In the \textbf{cold-start} scenario, SASRec demonstrates the weakest performance, confirming that CRMs struggle when historical user–item interactions are sparse. In contrast, both TALLRec and SASRecLLM perform significantly better, indicating that LLM-based methods can compensate for this limitation by leveraging world knowledge and contextual reasoning. Notably, SASRecLLM outperforms TALLRec across all evaluation metrics, highlighting the strength of the integrated collaborative modelling architecture and the effectiveness of the training strategies.

In the \textbf{warm-start scenario}, TALLRec achieves the lowest performance, suggesting that even when fine-tuned, relying solely on the LLM is insufficient despite the availability of rich interaction histories. This underscores the importance of incorporating collaborative modeling in such contexts. Interestingly, SASRec slightly outperforms SASRecLLM, implying that further improvements in collaborative modeling mechanisms and fine-tuning hyperparameters could enhance SASRecLLM’s performance in warm-start conditions.

Overall, the proposed framework effectively combines the strengths of collaborative recommendation and LLM-based reasoning. SASRec extends collaborative modeling with self-attentive sequential learning, enabling strong performance in warm-start scenarios, while the LLM component enhances generalization and alleviates the cold-start challenge through its contextual understanding and natural language capabilities.

\begin{figure*}[ht] % Two column figure (notice the starred environment)
	\includegraphics[width=\linewidth]{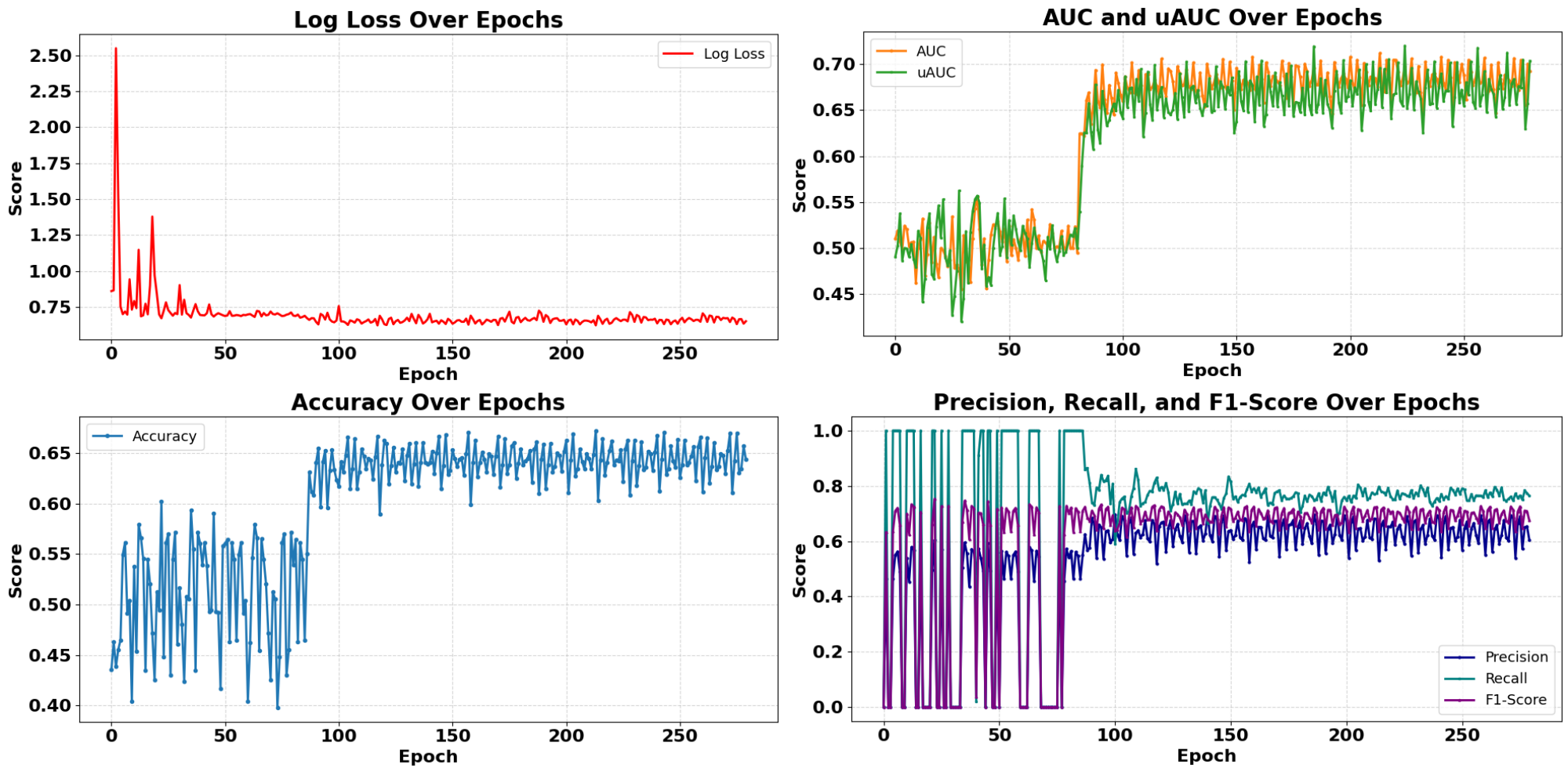}
    \caption{Training Log Line with Different Metrics in Movielens Dataset.}
    \label{fig: training_analysis}
\end{figure*}

\subsubsection{Training Analysis (RQ4)}

Fig. \ref{fig: training_analysis} illustrates SASRecLLM’s performance across training epochs in MovieLens dataset , showcasing the impact of the proposed training strategies over time. Due to the Dual-Stage Training strategy, SASRec is trained independently, and only the LoRA module is fine-tuned while the parameters of the mapping layer and SASRec remain frozen. The early training phase exhibits significant fluctuations, reflecting the model’s need for warm-up. Other evaluation metrics also remain low and unstable during this period. Once the early stopping criterion is triggered around epoch 80, signaling no further improvement in LoRA fine-tuning, the pre-trained SASRec is incorporated. At this point, the parameters of SASRec and the mapping layer are unfrozen, while the LoRA module is frozen. This transition leads to a noticeable improvement across all metrics, along with increased stability. These results validate the effectiveness and flexibility of the training strategies in optimizing SASRecLLM.

\section{Reflection}

Section \ref{sec: results} presents the experimental results of the proposed SASRecLLM framework, validating the effectiveness of combining fine-tuned LLMs with structured collaborative modeling. Experiments are conducted on two datasets using a diverse set of well-established baselines. Across all evaluation metrics, SASRecLLM demonstrates strong overall performance, confirming the efficacy of the proposed approach. To further assess the contribution of each component, an ablation study is performed using two SASRecLLM variants. As suggested by previous work \cite{zhao2023recommender, bao2023tallrec}, integrating LLMs into recommendation significantly boosts performance. However, the relatively modest performance gap between LLM4Rec models with and without collaborative modeling highlights the need for deeper investigation into the role of structured user-item interactions. Additionally, prior research has consistently shown that collaborative methods struggle in cold-start scenarios \cite{gogna2015comprehensive}, while LLM4Rec mitigate this issue by leveraging world knowledge and generalization capabilities. Therefore, SASRecLLM is evaluated in both warm and cold settings. The results satisfy the initial objectives: CRMs perform better when historical interactions are abundant, while LLMs handle cold start more effectively. SASRecLLM outperforms baselines in both cases by combining the strengths of both paradigms. Finally, a training analysis is conducted to examine the impact of the proposed dual-stage strategy. Metric trends align with theoretical expectations, further supporting the robustness and practicality of the training design. Overall, the main contributions of this study are summaried as follows:

\noindent $\bullet$ This work reviews the latest work for LLM4Rec and highlights the research gap.

\noindent $\bullet$ Proposes SASRecLLM, which fine-tunes LLMs with LoRA for efficient recommendation adaptation. It integrates SASRec for collaborative encoding, aligns outputs via a mapping layer, and employs three training strategies for effective optimization.

\noindent $\bullet$ Conducts in-depth experiments and analysis across two datasets and multiple baselines under different conditions. Results demonstrate the effectiveness and generalizability of the proposed framework.

\subsection{Limitation}
The primary limitation of this study lies in computational resource constraints. Due to this limitation, the task setup employs a small-scale LLM, TinyLlama-1.1B, as the backbone, which notably affects the performance of ICL. To prevent memory overload, the Amazon Book dataset is downsampled, which may reduce data richness and adversely affect model performance, as discussed in Section \ref{sec: overall_com}. Furthermore, although the dataset was partitioned into training, validation, and test sets following standard practice, K-fold cross-validation was not performed due to resource limitations. This may impact the robustness of the training process. The computational limitational also makes the training and evaluating time very long. Moreover, hyperparameter tuning is not applied during the eval-phase while training, which may have constrained the overall performance of SASRecLLM. Finally, all baselines are re-implemented and customized specifically for this study to maintain fairness and consistency. However, as baseline implementation and optimization were not the primary focus, slight performance deviations from results reported in prior work may occur.

\subsection{Future Work}
Building on the limitations, future work should begin by scaling the computational environment, specifically upgrading GPU and memory resources. This enhancement would support the use of larger language models and more comprehensive datasets, enabling the adoption of advanced training strategies and higher-fidelity evaluations. With improved infrastructure, the current binary classification setup could also be expanded to include more diverse recommendation tasks such as top-K ranking or multi-class classification, better leveraging SASRecLLM’s self-attentive sequential modeling capabilities. Additionally, future research should aim to improve baseline implementations and include a broader set of evaluation metrics, particularly those suited for SRSs to facilitate a more comprehensive and fair comparison. Lastly, further investigation into the cross-domain generalization and natural language understanding capabilities of LLMs would extend this work’s relevance across broader recommendation contexts.

Additionally, enhancing the modularity of SASRecLLM by integrating different LLM backbones and CRM encoders could further improve adaptability and flexibility. Exploring ensemble approaches that combine different modules from the ablation study (e.g., SASRec, TALLRec, and SASRecLLM) may also yield performance gains. Moreover, while this study employs a mapping layer to align collaborative embeddings with the LLM token space, alternative integration strategies such as cross-attention mechanisms or prompt engineering represent promising directions for future research. Although this study uses CRMs as the encoder, further exploration into alternative methods of embedding collaborative modeling for LLM4Rec remains an innovative area of inquiry.

\subsection{Conclusion}
This study proposes a novel framework, SASRecLLM, that integrates LLM4Rec through collaborative modeling and parameter-efficient fine-tuning. SASRec serves as the collaborative encoder using self-attention to model sequential interactions, while the LLM is fine-tuned via LoRA. A mapping layer aligns their embeddings for seamless integration. Empirical results demonstrate that SASRecLLM outperforms strong baselines across two datasets and excels in both cold-start and warm-start scenarios, validating its robustness and generalizability. Ablation studies further confirm the complementary contributions of the SASRec encoder and the fine-tuned LLM, underscoring the effectiveness of the combined architecture. This work contributes to the growing field of LLM4Rec by introducing a flexible and modular framework that unites structured CF with language-based reasoning. Future research could explore more scalable architectures, diverse CRMs encoders, and alternative integration mechanisms. Moreover, exploring the other roles that the CRM can play beyond encoder.

%----------------------------------------------------------------------------------------
%	 REFERENCES
%----------------------------------------------------------------------------------------

\printbibliography % Output the bibliography

%----------------------------------------------------------------------------------------

\end{document}